\documentclass[aps,pra,superscriptaddress,showpacs,twocolumn,floatfix]{revtex4-1}

\usepackage[T1]{fontenc}
\usepackage[utf8]{inputenc} 
\usepackage[english]{babel}   

\usepackage{graphicx}
\usepackage{hyperref}
\hypersetup{colorlinks=true,allcolors=blue}
\usepackage{xcolor}
\usepackage{physics}
\usepackage{amsfonts}
\usepackage{amsmath}
\usepackage{amssymb}
\usepackage{mathtools}

\usepackage{microtype}

\usepackage{ulem}
\newcommand{\e}[1]{\mathrm{e}^{#1}}

\begin{document}
\title{Low-energy prethermal phase and crossover to thermalization in nonlinear kicked rotors}
\author{Maxime Martinez}
\affiliation{Laboratoire  Kastler  Brossel,  Sorbonne  Universit\'e,  CNRS, ENS-PSL  University,  Coll\`ege  de  France,  4  Place  Jussieu,  75005  Paris,  France}
\author{Pierre-\'Elie Larr\'e}
\email{pierre-elie.larre@inphyni.cnrs.fr}
\affiliation{Universit\'e C\^ote d'Azur, CNRS, INPHYNI, France}
\author{Dominique Delande}
\email{dominique.delande@lkb.upmc.fr}
\affiliation{Laboratoire  Kastler  Brossel,  Sorbonne  Universit\'e,  CNRS, ENS-PSL  University,  Coll\`ege  de  France,  4  Place  Jussieu,  75005  Paris,  France}
\author{Nicolas Cherroret}
\email{nicolas.cherroret@lkb.upmc.fr}
\affiliation{Laboratoire  Kastler  Brossel,  Sorbonne  Universit\'e,  CNRS, ENS-PSL  University,  Coll\`ege  de  France,  4  Place  Jussieu,  75005  Paris,  France}
\begin{abstract}
In the presence of interactions, periodically-driven quantum systems generically thermalize to an infinite-temperature state. Recently, however, it was shown that in random kicked rotors with local interactions, this long-time equilibrium could be strongly delayed by operating in a regime of weakly fluctuating random phases, leading to the emergence of a metastable  thermal ensemble. Here we show that when the random kinetic energy is smaller than the interaction energy, this system in fact exhibits a much richer dynamical phase diagram, which includes a low-energy pre-thermal phase characterized by a light-cone spreading of correlations in momentum space. We develop a hydrodynamic theory of this phase and find a very good agreement with exact numerical simulations. We finally explore the full dynamical phase diagram of the system and find that the transition toward full thermalization is characterized by relatively sharp crossovers.
\end{abstract}
\maketitle

\section{Introduction}

When brought out of equilibrium, isolated quantum many-body systems generically involve a \textit{thermalization} process where local observables can be described by a Gibbs ensemble at sufficiently long time~\cite{Polkovnikov2011, Gogolin2016, Alessio2016, Deutsch2018}. Under specific circumstances, the out-of-equilibrium dynamics following a quantum quench can also exhibit a transient  \textit{pre-thermal} stage, where the system experiences dephasing  associated with the propagation of nearly-independent quasi-particles of very long lifetime~\cite{Berges2004, Kitagawa2011, Kollar2011, Buchhold2016, Larre2018, Mori2018, Martone2018, Mallayya2019}. In this case, the system truly thermalizes over a much longer time scale controlled by the collisions between quasi-particles. Such long-lived pre-thermal states have been observed in cold-atom~\cite{Gring2012, Langen2013, Langen2015} and photon-fluid~\cite{Abuzarli2022} experiments.

In out-of-equilibrium physics, periodically-driven interacting systems play a peculiar role due to the absence of energy conservation. While, generically, the interplay between driving and interactions makes the system evolve toward an infinite-temperature state~\cite{Reitter2017, Dalessio2014, Ponte2015b}, recently different scenarios have been  put forward.
Examples include the phenomenon of many-body dynamical localization, which brings the driven system to a stationary state~\cite{Ponte2015, Ponte2015b, Lazarides2015, Keser2016, Bordia2017, Notarnicola2020, Rylands2020, Vuatelet2021}, or the use of high-frequency driving or long-range interactions to induce metastable long-lived states~\cite{Abanin2015, Bukov2015, Else2017, Howell2019, Abadal2020, Hodson2021, Bhakuni2021}.

Among periodically-driven systems, the quantum kicked rotor has played a major role, in particular due to the phenomenon of dynamical localization~\cite{Casati1979}, a striking manifestation  of quantum interferences analogous to Anderson localization that has been thoroughly characterized experimentally~\cite{Moore95, Chabe2008, Hainaut17, Hainaut18, Hainaut2022}. In a gas of weakly interacting kicked bosons described at the mean-field level---the nonlinear kicked rotor (NKR)---it was shown that dynamical localization breaks down~\cite{Shepelyansky1993, Gligoric2011, Cherroret14, Cherroret16, Lellouch2020}, with the kinetic energy growing sub-diffusively up to arbitrarily long time. Recently, however, it was suggested that by operating in a regime where the strength of the random phases  is smaller than a certain threshold, one could inhibit inter-band transitions responsible for heating in the NKR, thus inducing a metastable state characterized by a thermal Gibbs ensemble~\cite{Haldar2022}.

In this article, we push this idea further and theoretically show  that in situations where the interaction strength becomes stronger than the random phases fluctuations, the NKR not only displays a thermal phase, but  also a {low-energy pre-thermal} phase for which we develop an analytical, hydrodynamic description.
Importantly, unlike the thermalization process discussed in~\cite{Haldar2022}, which stems from inelastic collisions between massive particles, the pre-thermal regime that we identify is built upon long-lived independent phononic excitations which make the system resemble  a superfluid at equilibrium \cite{Mu2022}. These excitations arise through the growth of exponential momentum correlations spreading within a light cone, a phenomenon that we study both theoretically and numerically. We also point out that in the pre-thermal phase, the NKR can be seen as the reciprocal version (in momentum space) of a weakly interacting, spatially disordered Bose gas of finite mean velocity in the low-energy limit~\cite{Scoquart2020b, Cherroret2021}, with the velocity being controllable via the phase of the kick modulation.
We finally construct the full non-equilibrium phase diagram of the system and, in particular, describe the crossover from the pre-thermal to the thermal phase and analyze how it is impacted by a change of the system's parameters.

The article is organized as follows. First, in Secs.~\ref{sec:analytics} and \ref{Sec:Pre-thermal dynamics}, we present a detailed analytical description of the low-energy phase in the NKR, based on an adaptation of the Bogoliubov-Popov theory of quantum fluctuations to classically fluctuating low-dimensional disordered systems. Then, in Sec.~\ref{Sec:numerics}, we compare our analytical findings with direct simulations of the NKR. We find an excellent agreement without any fitting parameter. In Sec.~\ref{Sec:crossover_therm}, we numerically explore the crossover from the pre-thermal to the thermal phase in the NKR, in particular making contact with the results reported in Ref.~\cite{Haldar2022},  and compute the full non-equilibrium phase diagram of the system. Section~\ref{Sec:conclusion} finally summarizes our findings. Technical details are collected in the Appendix.  

\section{Low-energy hydrodynamic theory of the nonlinear kicked rotor}

\label{sec:analytics}

\subsection{The nonlinear kicked rotor}
\label{SubSec:NKR}

We consider an assembly of $N_a$ weakly interacting bosons of mass $m$ on a ring of length $2\pi/k_r$, subjected to a periodically-kicked potential. Following~\cite{Shepelyansky1993, Lellouch2020, Haldar2022}, we consider a cubic local interaction with strength $g>0$ in momentum space 
\footnote{Note that the pre-thermalization scenario studied in Sec.~\ref{Sec:Pre-thermal dynamics} occurs regardless the structure of the interaction term in the nonlinear wave equation (\ref{eq:GPE}), be it local or nonlocal, in the latter case at least provided the response function varies rapidly at the scale of the healing length.} 
and model the dynamics using the Gross-Pitaevskii-type equation
\begin{align}
i\hbar\partial_t\psi={H}(t)\psi +g N_a |\psi|^2\psi
\label{eq:GPE}
\end{align}
for the wave function $\psi=\psi(k,t)$ in momentum space ($k$ is the wave number), with the time-dependent kicked-rotor Hamiltonian
\begin{equation}
H(t)= \frac{\hbar^2 {k}^2}{2m} - K\cos (k_r{x}  - \phi)  \sum_{n=0}^{\infty} \delta\Big(\frac{t}{T}-n\Big).
\label{eq:kicked-ham}
\end{equation}
Here $x\in[-\pi/k_r,\pi/k_r[$ is the position on the ring.
The second term in the right-hand side of~\eqref{eq:kicked-ham} describes a kicked cosine potential: It is switched on at every period $T$  and its amplitude $-K\cos(k_r{x}-\phi)$ depends on the position on the ring, $K>0$ being called the kicking strength.
Notice that  we have included a finite phase-shift $\phi$ in the cosine modulation (we choose $\phi\in [-\pi/2,\pi/2[$ in the following). Within the low-energy hydrodynamic mapping that one can construct from~\eqref{eq:kicked-ham} (see Sec.~\ref{Sec:LinearizationOfDynamicalEquations}), we will see that this parameter plays a role similar to a global velocity for a disordered one-dimensional Bose gas (see Sec.~\ref{Sec:analogy}). The wave function has the normalization (see Appendix~\ref{App:Conventions} for a summary of the conventions we use)
\begin{align}
    k_r \sum_k \vqty{\psi(k,t)}^2=\int_{-{\pi}/{k_r}}^{{\pi}/{k_r}} \frac{\dd{x}}{2\pi} \vqty{\psi(x,t)}^2 =1,
    \label{eq:normalization}
\end{align}
where the sum in reciprocal space runs over discrete wave numbers $k=l k_r$ ($l \in \mathbb{Z}$) because of the spatial periodicity of the Hamiltonian~\eqref{eq:kicked-ham}.

In the following, we study the time evolution of an initial \emph{plane wave} in momentum space, that is,  $\smash{\psi(k,t=0)=\sqrt{{\rho_0}/{N_a}}}$. Here $\rho_0=N_a/(Nk_r)$ is the uniform density of the wave in momentum space, where $N$ is the number of momentum lattice points (in the thermodynamic limit, $N_a, N\to\infty$ with the ratio $N_a/N$ constant). The evolution operator between two consecutive kicks corresponding to Eq.~\eqref{eq:GPE} reads 
\begin{align}
{U}(t+T,t)= 
\e{-i[ \alpha(k) +gN_a |\psi({k},t)|^2] \frac{T}{\hbar} } \e{ i K\cos (k_r x - \phi) \frac{T}{\hbar}},
\label{eq:propagator}
\end{align}
where $\alpha(k)=\hbar^2k^2/(2m)$ denotes the kinetic energy.

\subsection{Low-energy regime}
\label{Sec:lowenergy_regime}

A well-known regime of the \textit{non-interacting} quantum kicked rotor [$g=0$ in~Eq.~\eqref{eq:GPE}] corresponds to the limit of large kick amplitude $K$. 
In that case, after a kick a particle typically moves over a large distance and thus ends up at a completely different position on the ring, which strongly modifies the amplitude and the sign of the next kick. At long enough time, the particle is thus subjected to a series of kicks of quasi-random amplitudes, making the wave number $k$ a quasi-random variable. At large $K$ the non-interacting kicked rotor can thus be seen as a tight-binding model where a particle hops between  sites of random momentum as a result of the kicks, which is the counterpart in momentum space of an Anderson disorder model in position space~\cite{Fishman1982,Grempel84,Shepelyansky1986}. In this picture, the kinetic phases $\alpha(k)T/\hbar$ in Eq.~\eqref{eq:propagator} play the role of the onsite disorder. As such, they are often taken as random numbers evenly distributed in the interval $[0,2\pi[$ (provided $\hbar k_r^2T/(4 m \pi)$ is irrational; in the opposite case, quantum resonances occur~\cite{Izrailev1980,Sokolov2000,Wimberger2004,Lepers2008}).

In the present article, however, we consider a different, low-energy regime 
where both the kinetic phases and the phases induced by the kicks are small compared to $2\pi$. Denoting by $W$ the typical fluctuations of $\alpha(k)$, this condition reads:

\begin{equation}
\frac{KT}{\hbar},\, \frac{WT}{\hbar}\ll2\pi.
\label{cond_KW}
\end{equation}

In a practical experiment, a weak value of $W$ might be achieved by operating in the close vicinity of a quantum resonance. 
In addition, in the following we will focus on a regime where the interaction strength much \emph{exceeds} the  fluctuations of the kinetic energy:
\begin{equation}
\frac{W}{g\rho_0}\ll 1.
\label{eq:lowenergy}
\end{equation}
Together with (\ref{cond_KW}), this condition guarantees
that the dynamics of the kicked particles becomes essentially dominated by low-lying Bogoliubov phonons, yielding an enhanced coherence of the system. In the recent work~\cite{Haldar2022}, typical values for $W$ and $g\rho_{0}$ were such that $W/(g\rho_0) \simeq 8$, implying that the observed dynamics of the NKR was mainly governed by disorder scattering events on top of which (rare) inelastic collisions  were slowly thermalizing the system, similarly to previous works considering disorder in position space~\cite{Cherroret2015, Scoquart2020}. In strong contrast, when the inequality~\eqref{eq:lowenergy} is satisfied  the density  fluctuations of the wave function become strongly suppressed, corresponding to a suppression of particle scattering, and a pre-thermal phase can emerge.

\subsection{Hydrodynamic equations}

\label{Sec:LinearizationOfDynamicalEquations}

To describe the low-energy phase in the NKR, we start by expressing the stroboscopic evolution of the wave function over one period. Using the momentum-space representation of the operator $\exp[iK\cos(k_r x-\phi)T/\hbar]$ and the fact that the kinetic and interaction energies are local in momentum space, Eq.~\eqref{eq:propagator} gives us 
\begin{align}
\notag
\psi(k,t+T)&\left.=\exp\!\bigg\{{-}i [\alpha(k)+g N_a |\psi(k,t)|^2] \frac{T}{\hbar} \bigg\}\right. \\
\label{eq:exact-dyn-ext}
&\left.\hphantom{=}\times\sum_{l=-\infty}^{\infty} i^l \e{-il \phi} J_l\qty(\frac{K T}{\hbar})\psi\qty(k+lk_r,t).\right.
\end{align}
Under the conditions of weak kinetic phases and weak kick amplitudes introduced in Sec.~\ref{Sec:lowenergy_regime}, the wave function tends to retain a robust coherence in momentum space. In other words, $\psi(k,t)$ becomes a weakly varying function of $k$. This allows us to linearize the wave function as
\begin{align}
\psi\qty(k+l k_r,t) \simeq \psi(k,t) +  lk_r\partial_k\psi + \dfrac{l^2k_r^2 }{2}\partial_k^2\psi.
\label{eq:momentum_expansion}
\end{align}
Note that this expansion assumes a continuous approximation of the discrete wave-vector-$k$ basis, whose relevance will be  discussed in Sec.~\ref{Sec:validity}. 
We then insert Eq.~\eqref{eq:momentum_expansion} in the second line of Eq.~\eqref{eq:exact-dyn-ext}, and use the identities
\begin{subequations}
\begin{align}
\sum_{l=-\infty}^{\infty} i^l \e{-il \phi} J_l\qty(\frac{K T}{\hbar})&=e^{i \frac{K T}{\hbar} \cos \phi},\\
\sum_{l=-\infty}^{\infty} li^l \e{-il \phi} J_l\qty(\frac{K T}{\hbar})&=\frac{KT}{\hbar}\ \sin\phi\ e^{i \frac{K T}{\hbar} \cos \phi},\\
\sum_{l=-\infty}^{\infty} l^2i^l \e{-il \phi} J_l\qty(\frac{K T}{\hbar})&\simeq \frac{iKT}{\hbar}\cos\phi\, e^{i \frac{K T}{\hbar} \cos \phi}. \label{sum_square}
\end{align}
\end{subequations}
In Eq.~\eqref{sum_square}, we have dropped a quadratic correction $(KT/\hbar)^2\sin^2\phi\ e^{i \frac{K T}{\hbar} \cos \phi}$, given that the kick strength is sufficiently small, see Eq.~(\ref{cond_KW}).
	Equation~\eqref{eq:exact-dyn-ext} becomes
\begin{align}
\label{eq:propagation-simpl}
&\left.\psi(k,t+T)\simeq \right. \\
&\left.\quad\exp\!\bigg\{{-}i [\alpha(k)\!+\!gN_a|\psi(k,t)|^2 \! -\!  K \cos \phi] \frac{T}{\hbar}\bigg\}\nonumber\right.\\
&\left.\quad\times \bigg[\psi(k,t) +  \frac{K k_r T\sin \phi }{\hbar}\,   \partial_k\psi+ i    \frac{K k_r^2 T\cos \phi}{2 \hbar}   \partial^2_k\psi\bigg].\right. \notag
\end{align}

Following a standard procedure for treating low-dimensional Bose gases~\cite{Popov1972, Popov1983, Mora2003}, we start by expressing the wave function in the polar form (known as ``Madelung transformation'')
\begin{equation}
    \psi(k,t)= \sqrt{\frac{\rho(k,t)}{N_a}}~
    \!\exp\!\Big[i \theta(k,t)\!-\! i(g \rho_0 \!-\! K \cos  \phi) \frac{t}{\hbar}\Big],
    \label{eq:def_density_phase}
\end{equation}
where $\rho(k,t)$ and $\theta(k,t)$ are the system's density and phase in momentum space. The gauge factor $\exp[i(g \rho_0 \!-\! K \cos  \phi)t/\hbar]$ is introduced here for convenience, as it allows one to eliminate constant corrections in the equations of motion below. In the spirit of Eq.~\eqref{eq:momentum_expansion}, we then assume that the wave function varies weakly in time during a period, so that
\begin{align}
    \psi(k,t+T) \simeq \psi(k,t) + T \partial_t\psi,
    \label{eq:time_expansion}
\end{align}
and we write the momentum-space density  $\rho(k,t)=\rho_0+\delta \rho(k,t)$ in terms of its fluctuations on top of the uniform background $\rho_0$. Combining Eqs.~\eqref{eq:propagation-simpl}--\eqref{eq:time_expansion}, we obtain
\begin{align}
\dfrac{1}{2\sqrt{\rho}}
\partial_t \delta \rho 
 & +i\sqrt{\rho}\,
\partial_t  \theta 
  = \dfrac{\sqrt{\rho}}{T}
  \Big[\e{-i\qty(\alpha(k) +g \delta \rho)\frac{T}{\hbar}}-1\Big]\nonumber\\
&+ \!  \dfrac{K k_r \sin \phi}{\hbar}   \e{-i\qty(\alpha(k) +g \delta \rho)\frac{T}{\hbar}} \qty[\dfrac{1}{2\sqrt{\rho}} 
\partial_k\delta \rho \!+\!i\sqrt{\rho}\partial_k\theta]\nonumber \\
& + i \dfrac{K k_r^2 \cos \phi}{2\hbar}  \e{-i\qty(\alpha(k) +g \delta \rho) \frac{T}{\hbar}}\nonumber\\ 
&\times\qty[\partial_k^2\sqrt{\rho} 
+ \dfrac{i}{\sqrt{\rho}}\partial_k\qty(\rho\, \partial_k\theta)
- \sqrt{\rho}\qty(\partial_k\theta)^2]. 
\label{eq:expanded_full}
\end{align}

To simplify the nonlinear hydrodynamic equation~\eqref{eq:expanded_full}, we expand it with respect to $\alpha$ using Eq. (\ref{cond_KW}), as well as with respect to the density and phase-gradient fluctuations, $\delta\rho$ and $\partial_k\theta$, respectively.  The latter expansion stems from the condition (\ref{eq:lowenergy}) and will be justified a posteriori in Sec.~\ref{Sec:lowenergy_regime}.

Equating the real and imaginary part of this expansion, we end up with the following coupled Bogoliubov-de Gennes-type equations for the fluctuations of the NKR:
\begin{align}
 &\partial_t\delta \rho\!=\!   \frac{ K k_r \sin \phi}{\hbar}  \partial_k \delta \rho -   
 \rho_0  \frac{K k_r^2 \cos \phi}{\hbar} ~  \partial_k^2\theta, \label{eq:bog_kr_rho}\\
 & \partial_t\theta \!= \! \frac{ K k_r \sin \phi}{\hbar}  \partial_k\theta\!  +\!  \frac{ K k_r^2 \cos \phi}{ 4 \hbar \rho_0} \partial_k^2\delta \rho\! -\! \frac{\alpha(k)}{\hbar}\!-\! \frac{g \delta \rho}{\hbar}.\label{eq:bog_kr_theta}
\end{align}

\subsection{Analogy with a Bose gas moving in a disorder potential}
\label{Sec:analogy}

Before examining the solutions of Eqs.~\eqref{eq:bog_kr_rho} and~\eqref{eq:bog_kr_theta}, it is interesting to notice that they are analogous to the dynamical equations that govern the density and phase fluctuations of a quasi-one-dimensional atomic Bose-Einstein condensate moving at a certain velocity $-v<0$ in a spatially random potential $V(x)$ (a two-dimensional version of this problem has been studied in~\cite{Scoquart2020b, Cherroret2021}). For such a system, the Gross-Pitaevskii equation for the order parameter $\psi=\psi(x,t)$ reads
\begin{equation}
    i\hbar\partial_t\psi=H\psi+g N_a|\psi|^2\psi,
    \label{GPE_disorder}
\end{equation}
where
\begin{align}
H = -\frac{\hbar^2}{2m}\partial_x^2 + V(x) +i v \hbar \partial_x
\end{align}
is the Hamiltonian without interactions in the comoving frame. Looking for a solution of the form $
\psi(x,t)=\sqrt{\rho(x,t)} \e{i \theta(x,t) -i{g \rho_0}t/{\hbar}}$ with $\rho(x,t)=\rho_0+\delta \rho(x,t)$, and expanding Eq.~\eqref{GPE_disorder} to first order in the disorder potential $V(x)$ and in the fluctuations it induces (linear-response approach), one finds~\cite{Scoquart2020b, Cherroret2021}
\begin{align}
\partial_t \delta \rho &= v  \partial_x \delta \rho
-\frac{\hbar \rho_0 }{m}
\partial_x^2\theta, \\
\partial_t\theta&= v  
\partial_x\theta+ \dfrac{\hbar}{4m \rho_0}
\partial_x^2\delta\rho - \dfrac{V(x)}{\hbar} - \dfrac{g \delta \rho}{\hbar},
\end{align}
whose analogy with Eq.~\eqref{eq:bog_kr_theta} is transparent. In particular, we have the following correspondences in the NKR:
\begin{align}
x&\longleftrightarrow k
&V(x)&\longleftrightarrow \alpha(k)
\nonumber\\
v&\longleftrightarrow \frac{K k_r\sin\phi}{\hbar} 
 &m&\longleftrightarrow \frac{\hbar^2}{k_r^2K\cos\phi}.\nonumber 
\end{align}
Observe, in particular, that changing the parameter $\phi$---which originally appeared as a phase shift in the modulation amplitude of the kicks in Eq.~\eqref{eq:kicked-ham}---amounts to modifying the mean gas velocity and the effective mass in the position-space mapping. As long as $|\phi|$ is not too close to $\pi/2$, however, the modification of the effective mass does not have any qualitative impact on the mapping.

\section{Pre-thermal dynamics}
\label{Sec:Pre-thermal dynamics}

\subsection{General solution of the Bogoliubov-de Gennes equations}

For the initial plane-wave state $\psi(k,t=0)=\sqrt{{\rho_0}/{N_a}}$, the initial values of the density fluctuations and of the phase are
\begin{align}
    \delta \rho(k,t=0)=0\quad\text{and}\quad\theta(k,t=0)=0.
    \label{eq:initial_condition} 
\end{align}

Equipped with these initial conditions, we solve the Bogoliubov-de-Gennes equations 
\eqref{eq:bog_kr_rho} and~\eqref{eq:bog_kr_theta} by introducing the new variables $\varphi_1=\delta \rho/\sqrt{\rho_0}$ and $\varphi_2=2i\sqrt{\rho_0} \theta$ and their Fourier transform
\begin{align}
    \tilde{\varphi}_i(x,t) &= k_r \sum_{k} \varphi_i(k,t)  \e{i kx }, \\
    \varphi_i(k,t) &=   \int_{-{\pi}/{k_r}}^{{\pi}/{k_r}} \frac{\dd{x}}{2\pi}  \tilde{\varphi}_i(x,t)  \e{i k x},
    \label{eq:fourier_transform}
\end{align}
where the sum runs over discrete $k= l k_r$ with integer $l$ (see Appendix~\ref{App:Conventions} for a summary of the conventions we use).
Going to Fourier space allows us to rewrite~Eqs.~\eqref{eq:bog_kr_rho} and~\eqref{eq:bog_kr_theta} as the linear system  \begin{gather}
i\hbar \partial_t \mqty(\tilde{\varphi}_1\\\tilde{\varphi}_2) = 
\mathcal{M}\mqty(\tilde{\varphi}_1\\\tilde{\varphi}_2) + 2\sqrt{\rho_0} \tilde{\alpha} (x) \mqty(0 \\ 1),
\label{eq:sys_phi12}
\end{gather}
where $\tilde{\alpha}(x)$ is the Fourier transform~\eqref{eq:fourier_transform} of the random phases $\alpha(k)$, and 
\begin{equation}
    \mathcal{M}=\mqty(v \hbar x & \epsilon_x  \\
\epsilon_x+2g\rho_0  & v \hbar x)
\end{equation}
is the Bogoliubov-de Gennes Hamiltonian in the density-phase representation, with $\epsilon_x=\frac{1}{2} k_r^2 x^2 K \cos \phi $ and $v=K k_r\sin\phi/\hbar$ (see Sec.~\ref{Sec:analogy}).

The solution of the linear system~\eqref{eq:sys_phi12} with initial conditions~\eqref{eq:initial_condition} is formally given by
\begin{gather}
\mqty(\tilde{\varphi}_1\\\tilde{\varphi}_2) = -2i\sqrt{\rho_0} \frac{\tilde{\alpha}(x)}{\hbar} \int_0^t \dd{t'} 
\e{i\frac{t'-t}{\hbar}\mathcal{M}}
\mqty(0 \\ 1).
\label{eq:sol_formal}
\end{gather}
To compute the matrix exponential, we diagonalize $\mathcal{M}$.
Its two eigenvectors are $U_\pm=(\pm u_1,u_2)^T$, where
\begin{align}
    u_1 = \sqrt{\frac{\epsilon_x}{\epsilon_x+2g\rho_0}}  \qqtext{and} u_2= \sqrt{\frac{2g\rho_0 }{\epsilon_x+2g\rho_0}}.\label{eq:eigenvec_M}
\end{align}
The corresponding eigenvalues are $\Lambda_\pm = v \hbar x \pm E_x$, where $E_x=\sqrt{\epsilon_x\qty(\epsilon_x+2g \rho_0)}$ is the Bogoliubov spectrum for the system at rest. The latter is quadratic at large $x$ and becomes linear at small $x$, $E_{x}\simeq c_{s}\hbar |x|$, where
\begin{equation}
\label{Eq:BogSound}
    c_s=k_r \sqrt{\frac{g \rho_0 K \cos \phi }{\hbar^2}}
\end{equation}
is the Bogoliubov speed of sound.
We also define the healing length of the system, $\xi$, as the typical scale (in $k$ space) separating these large- and low-$x$ regimes, that is, $\epsilon_x \sim 2g\rho_0$ typically for $x\sim 2/\xi$. This gives
\begin{align}
\label{eq:def_xi}
    \xi = k_r \sqrt{\frac{K \cos \phi}{ g \rho_0}}.
\end{align}

The exponential of $\mathcal{M}$ is now diagonal in the basis of $U_\pm$ and can readily be expressed in the original basis, using the change-of-basis matrix from Eq.~\eqref{eq:eigenvec_M}. Performing the left integration in Eq.~\eqref{eq:sol_formal} and coming back to the initial variables, we finally obtain
\begin{widetext}
\begin{align}
\delta \rho(k,t) &= 2\rho_0 \int_{-\pi/k_r}^{\pi/k_r} \frac{\dd{x}}{2\pi} \tilde{\alpha}(x) \frac{\epsilon_x}{E_x} \frac{E_x [\cos (xv t) - \cos(E_x t/\hbar)] - i E_x \sin (xvt)+ i v \hbar x \sin (E_x t/\hbar)}{v^{2} \hbar^{2} x^{2}- E_x^{2}}\e{-i x(k-vt)} , \label{eq:bog-theta-rho-result_deltarho}  \\ 
\theta (k,t) &=  \int_{-\pi/k_r}^{\pi/k_r} \frac{\dd{x}}{2\pi} \tilde{\alpha}(x) \frac{v \hbar x \sin(xv t)- E_x\sin(E_x t/\hbar)+i v \hbar x [ \cos (xv t)- \cos (E_x t/\hbar)] }{v^{2} \hbar^{2} x^{2}- E_x^{2}}\e{-i x(k-vt)} . \label{eq:bog-theta-rho-result_theta} 
\end{align}
\end{widetext}

\subsection{Coherence function}
\label{subsec:coherence}

To exemplify the above formalism, we compute the same-time two-field correlation function of the system in momentum space, which describes the time evolution of the spatial coherence of the Bose gas:
\begin{gather}
g_1(\Delta k,t) = \frac{  \overline{\psi^*(0,t) \psi(\Delta k,t)}}{{\vqty{\psi(0,t)}^2}} ,
\label{g1_def}
\end{gather}
where the overbar refers to an ensemble average over the random energies $\alpha$, and
 $\overline{|\psi(0,t)|^2}=\rho_0/N_a$ is the density of the initial plane wave.
To evaluate this correlator, we insert Eq.~\eqref{eq:def_density_phase} into Eq. (\ref{g1_def}) and neglect the density fluctuations, i.e., we use $\rho(k,t)=\rho_0+\delta\rho(k,t)\simeq \rho_0$. This approximation is motivated by the well-known property that  phase fluctuations in general dominate over density fluctuations 
in low-dimensional Bose systems at weak interactions \cite{Mora2003, Petrov2003}. In the dynamical problem considered here, this property becomes satisfied very quickly, typically after an evolution time $\sim \hbar/(g\rho_0)$ \footnote{The way the subleading density fluctuations $\delta\rho(k,t)$ contribute to the $g_{1}$ function may be found in, e.g., Ref.~\cite{Larre2018} [see Eq.~(84)]}. 
Equation~\eqref{eq:def_density_phase} then leads to 
$g_1(\Delta k,t) \simeq\overline{\exp\{i[\theta(0,t)-\theta(\Delta k,t)]\}}$. 
Next we use that within the linearization procedure considered here, the Hamiltonian is quadratic so that the phase variance is a Gaussian random variable \cite{Mora2003}:
\begin{align}
g_1(\Delta k,t) \simeq \exp\!\Big\{{-}\frac{1}{2}\overline{\vqty{\theta(0,t)-\theta(\Delta k,t)}^2}\Big\}.
\label{eq:g1_fluct_theta}
\end{align}
We now assume that the correlations of the kinetic energies $\alpha(k)$ have a statistical translational symmetry, i.e., that their correlator takes the form $ \overline{\alpha^*(k_m) \alpha(k_n)}=\tilde{C}(k_n-k_m)$. It follows that \begin{align}
     \overline{\tilde{\alpha}^*(x) \tilde{\alpha}(x')}  = 2\pi C (x) \delta(x-x'),
     \label{eq:corrx}
\end{align}
where $C(x)=k_r\sum_l \tilde{C}(k_l)e^{ik_r l x}$ is the inverse Fourier transform of $\tilde{C}(k_n)$.
Inserting Eqs.~\eqref{eq:bog-theta-rho-result_theta} and~\eqref{eq:corrx} into Eq.~\eqref{eq:g1_fluct_theta}, we find, after some algebra,
\begin{align}
\ln g_1&(\Delta k,t) = 
\int \frac{\dd{x}}{2\pi} C(x)\sin^2\qty(\frac{\Delta k x}{2})\bigg\{
\frac{2\sin^2\qty({E_x t}/{\hbar})}{v^2 \hbar^2 x^2-E_x^2}\nonumber\\
&-\frac{4v\hbar x (v \hbar x+ E_x)  }{(v^2 \hbar ^2 x^2-E_x^2)^2} \sin^2\qty[\frac{(v \hbar x - E_x)t}{2 \hbar}] \nonumber \\
&-\frac{ 4v \hbar x (v \hbar x - E_x)}{(v^2 \hbar^2 x^2-E_x^2)^2}  \sin^2\qty[\frac{(v\hbar x +E_x)t}{2 \hbar}]\bigg\}.
\label{eq:g1-integrale}
\end{align}
From now on, we restrict ourselves to $\delta$-correlated kinetic energies $\alpha(k)$, corresponding to a uniform spectrum $C(x)=k_r W^2/12 $ [equivalently, $\tilde{C}(k_n-k_m)=(W^2/12) \delta_{nm}$]. The proportionality factor $1/12$ is chosen here so that $W^2/12$ coincides with the variance $\overline{\alpha(k)^2}$ of a uniform onsite distribution of the $\alpha$'s in the interval $[-W/2,W/2]$, which will be used in the numerical simulations of Sec.~\ref{Sec:numerics}. The integrals in Eq.~\eqref{eq:g1-integrale} range from $-\pi/k_r$ to $\pi/k_r$. However, at long enough time (typically, $t\gg\hbar/g\rho_0$), they are dominated by small $x$-values, so that these bounds can be extended to $\pm\infty$. Furthermore, in that limit the dispersion relation is accurately described by its phononic branch:  $E_x \simeq c_s\hbar |x|$.  

We first consider  Eq.~\eqref{eq:g1-integrale} for a vanishing effective velocity of the Bose gas, $v=K k_r\sin\phi/\hbar=0$ (i.e., $\phi=0$). We find
\begin{equation}
\label{eq:lightcone}
\ln g_1(\Delta k,t)\simeq
\begin{dcases}
- \dfrac{W^2k_r}{48 (g\rho_0)^2\xi}  \dfrac{|\Delta k|}{\xi},& |\Delta k|\ll 2c_st,\\
- \dfrac{W^2k_r}{48 (g\rho_0)^2\xi} \dfrac{ 2c_s t}{\xi},& |\Delta k|\gg 2c_st,
\end{dcases}
\end{equation}
which describes a light-cone spreading of the correlations at the Bogoliubov speed of sound $c_s$. More precisely, $g_1$ decays exponentially with $|\Delta k|$ up to the Lieb-Robinson bound $|\Delta k| = 2c_s t$, where it reaches a plateau whose height decays exponentially in time.
This behavior basically comes from the interference between quench-induced phonon excitations with momenta $\pm\hbar x$ and same energy $\hbar \omega = E_x \simeq c_s \hbar |x|$. Note that within the light cone, Eq.~\eqref{eq:lightcone} predicts a \emph{time-independent} coherence function. This is a characteristic feature of a pre-thermal dynamics~\cite{Berges2004, Kitagawa2011, Kollar2011, Buchhold2016, Larre2018, Mori2018, Martone2018, Mallayya2019}, where a non-equilibrium system at short time is governed by nearly-independent quasi-particles and exhibits an extremely slow dynamics. 

The integral in Eq.~\eqref{eq:g1-integrale} can also be evaluated in the general case where $v\ne 0$. In the subsonic regime $v<c_s$, the coherence function is governed by interference between phonons of Doppler-shifted energies $|E_x \pm v  \hbar x| \simeq (c_s\pm{v}) \hbar \vqty*{x}$. This gives rise to
four dynamical regimes depending on the value of $|\Delta k|$ compared to the three dynamical lengths $(c_s \pm v)t$ and $2c_s t$ (see Fig.~\ref{Tk_regimes}). In regions (I)--(IV), we find the following behaviors for the $g_{1}$ function:
\begin{figure}
    \centering
    \includegraphics{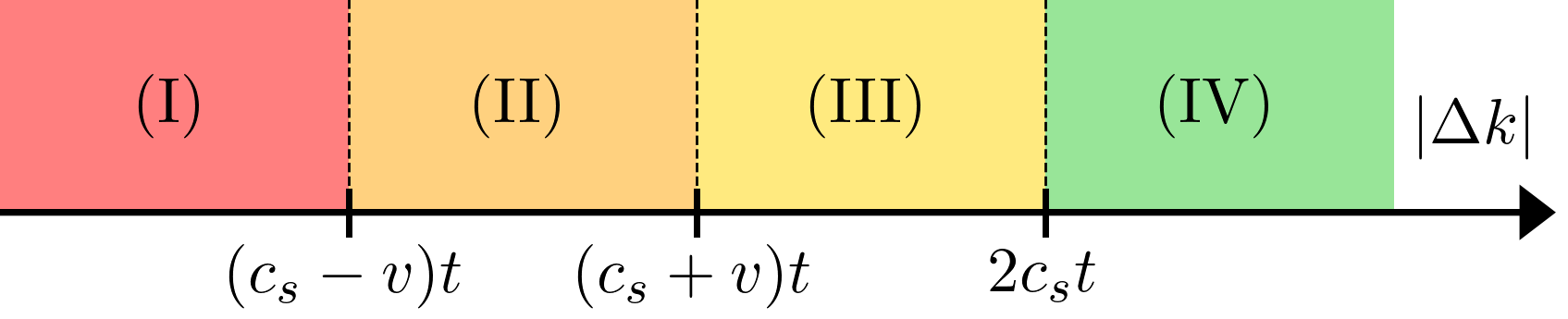}
    \caption{Four possible dynamical regimes of the coherence function $g_1$ when $v\ne 0$ and $v<c_s$. The expression of $g_1$ in each region is given by Eq.~\eqref{eq:g1-integrale-approx-2}.
    \label{Tk_regimes}}
\end{figure}
\begin{align}
\label{eq:g1-integrale-approx-2}
&\left.\ln g_{1}(\Delta k,t)\simeq\right. \\
\notag
&\left.
\begin{dcases}
-\frac{W^2k_{r}}{48(g\rho_{0})^{2}\xi}\frac{1+3v_{r}^{2}}{(1-v_{r}^{2})^{2}}\frac{|\Delta k|}{\xi} & \text{(I)}, \\
-\frac{W^2k_{r}}{48(g\rho_{0})^{2}\xi}\frac{1}{(1+v_{r})^{2}}\bigg(\frac{|\Delta k|}{\xi}+v_r\frac{1+v_{r}}{1-v_{r}}\frac{2c_{s}t}{\xi}\bigg)  & \text{(II)}, \\
-\frac{W^2k_{r}}{48(g\rho_{0})^{2}\xi}\frac{1}{1-v_{r}^{2}}\frac{|\Delta k|}{\xi} & \text{(III)}, \\
-\frac{W^2k_{r}}{48(g\rho_{0})^{2}\xi}\frac{1}{1-v_{r}^{2}}\frac{2c_{s}t}{\xi} & \text{(IV)},
\end{dcases}
\right.
\end{align}
where
\begin{equation}
	v_r = \frac{v}{c_s}
\end{equation}
is the incoming fluid velocity relative to the speed of sound \eqref{Eq:BogSound} (known as the Mach number). Equation~\eqref{eq:g1-integrale-approx-2} still describes an exponential decay of the coherence function  up to $|\Delta k| = 2c_s t$ with, however, two changes of slope at $|\Delta k| = (c_s \pm v) t$. For  $|\Delta k| > 2c_s t$, $g_1(\Delta k,t)$ still reaches a plateau whose height decays exponentially in time. 

Note that Eq.~\eqref{eq:g1-integrale} also admits a well-defined limit in the supersonic regime $v_r>1$. As discussed in the next section, however,  the validity of the approach is no longer guaranteed in that case. Furthermore,  under practical conditions the supersonic regime is not easily observable due to the onset of the thermal phase at relatively short time when $v_r>1$. This point will be discussed more in detail in Sec.~\ref{Sec:crossover_therm}.  

\subsection{Validity of the theory}
\label{Sec:validity}

Let us now discuss the range of validity of our hydrodynamic description of the NKR. 
One of the main assumptions used to derive the Bogoliubov-de Gennes equations~\eqref{eq:bog_kr_rho} and \eqref{eq:bog_kr_theta} is the smallness of the density fluctuation $\delta\rho$ compared to the background density $\rho_{0}$. To assess the validity of this approximation, we evaluate a posteriori the square root of $\overline{\delta\rho^{2}}/\rho_{0}^{2}$ from Eq.~\eqref{eq:bog-theta-rho-result_deltarho}. This ratio is at all times bounded from above by its long-time, $t\gg\hbar/(g\rho_{0})$ value, which for $v_{r}<1$ reads
\begin{equation}
\label{cond_smallrho}
\frac{\overline{\delta\rho^{2}}^{1/2}}{\rho_{0}}\simeq\frac{W}{g\rho_{0}}
\begin{dcases}
\frac{\sqrt{3+v_{r}^{2}}}{|1-v_{r}^{2}|}, & \xi\ll k_{r}, \\
\sqrt{\frac{k_{r}}{\xi}}\frac{\sqrt{3v_{r}^{2}\!-\!1\!+\!(1\!-\!v_{r}^{2})^{3/2}}}{v_{r}|1-v_{r}^{2}|^{3/4}}, & \xi\gg k_{r}.
\end{dcases}
\end{equation}
The small- and large-$\xi/k_{r}$ estimates~\eqref{cond_smallrho} show that sufficient conditions for the density fluctuations to be small are that $W/(g\rho_{0})\ll1$ and $v_{r}$ is not too close to the singular sonic limit $v_{r}=1$, at which the Bogoliubov phonons generated in the fluid have the tendency to pile up in the vicinity of the disorder grains, resulting in nonlinear fluctuations that cannot be captured by the present perturbative approach~\cite{Leboeuf2001, Larre2012}. In the supersonic regime $v_r>1$, we find from Eq.~\eqref{eq:bog-theta-rho-result_deltarho} that density fluctuations diverge and the approach is no longer reliable.
It is worth noting that in the most favorable case where the fluid is at rest ($v_{r}=0$), one recovers the necessary condition~\eqref{eq:lowenergy} discussed in Sec.~\ref{SubSec:NKR}~\cite{Scoquart2020, Cherroret2021}.

Second, we examine the validity of the assumption of weak variations of the wave function in momentum space used in the Taylor expansion~\eqref{eq:momentum_expansion}. To this aim, we note from Eq.~\eqref{eq:g1-integrale-approx-2} that the coherence function decays over the typical (momentum) scale $\delta k=(g\rho_{0}\xi)^{2} (1-v_r^2)^2/[W^{2}k_{r} (1+3v_r^2)]$. The expansion ~\eqref{eq:momentum_expansion} requires $\delta k\ll k_r$, which reads
\begin{equation}
\label{eq:cond_discrete}
\frac{W}{g\rho_{0}}\ll\frac{\xi}{k_{r}} \frac{1-v_r^2}{\sqrt{1+3v_r^2}}.
\end{equation}
Note, again, a breakdown of the approach when $v_r\sim 1$. In the case $v_r=0$, the right-hand side of this inequality reduces to $\xi/k_r$.  In the numerical simulations presented below we use nonlinearity and kick amplitudes such that this ratio is never far from unity, leading again to the condition~\eqref{eq:lowenergy}. 

\section{Numerical simulations in the pre-thermal phase}
\label{Sec:numerics}

\subsection{Numerical method}

We now compare our analytical predictions for the pre-thermal dynamics to numerical simulations. For these simulations, we set $\hbar = k_r = T =1$ and work with a finite system size $N$ for the momentum grid. Precisely, the wave numbers $k$ take the values  $k = -N/2+1,\dots 0, \dots N/2$ (for $N$ even). We also use periodic boundary conditions, such that the position $x$ also takes discrete values $x=\pm {\pi}/{N}, \pm {3 \pi}/{N}, \dots \pm {(N-1) \pi}/{N}$ and the normalization condition is written as
\begin{align}
\label{eq:normalization_num}
    \frac{1}{N} \sum_{x} \vqty{\psi(x,t)}^2 = \sum_k \vqty{\psi(k,t)}^2 =1,
\end{align}
with the Fourier transform relation 
\begin{align}
    \tilde{\varphi}_i(x) &=  \sum_{k} \varphi_i(k)  \e{i kx },  \label{eq:fourier_transform_discrete1}\\
    \varphi_i(k)&=   \frac{1}{N} \sum_x  \tilde{\varphi}_i(x)  \e{i k x}.
    \label{eq:fourier_transform_discrete2}
\end{align}
Note that we recover Eq.~\eqref{eq:fourier_transform} in the limit $N \rightarrow \infty$. Finally, we choose the initial plane-wave density $\rho_0=1$, that is, $N_a=N$.

To study the temporal evolution of the wave function, we use a split-step-like numerical scheme, using that the time-propagator~\eqref{eq:propagator} between two consecutive kicks is the product of two operators: $U(k,t+1)=U_k\times U_x $, with
\begin{align}
    U_x &=\e{i K \cos(x-\phi)}\\
    U_k &= \e{-i \alpha(k) -i gN_a \qty|\psi(k,t)|^2},
\end{align}
where $U_k$ is diagonal in the wave-vector basis and $U_x$ is diagonal in the position basis.
To propagate $\psi(k,t)$ to the next kick $\psi(k,t+1)$, we thus apply the following scheme
\begin{align}
    \psi(x,t+0^+) &= U_x \text{FFT}^{-1}[\psi(k,t)],\\
    \psi(k,t+1) &= U_p \text{FFT}[\psi(x,t+0^+)],
\end{align}
where $\psi(x,t+0^+)$ refers to  the wave function just after the kick $t$ and \text{FFT} represents the Fast Fourier Transform used as numerical implementation of Eqs.~\eqref{eq:fourier_transform_discrete1} and \eqref{eq:fourier_transform_discrete2}.

Finally, to compute $g_1(\Delta k,t)$ efficiently we also use the FFT and the relation~\eqref{eq:g1_fluct_theta}
\begin{align}
    g_1(\Delta k,t) = \text{FFT}^{-1}[\overline{\vqty{\psi(x,t)}^2}],\label{eq:g1_nx}
\end{align}
where the average is performed over $n_d$ realizations of the random phases $ \alpha(k)$, which we choose uncorrelated and uniformly distributed in the interval $[-W/2,W/2]$ (so that $\smash{\overline{\alpha(k)^2}=W^2/12}$). One may easily check that this corresponds to a power spectrum $C(x)=k_r {W^2}/{12} $.

\begin{figure*}[t]
    \centering
    \includegraphics{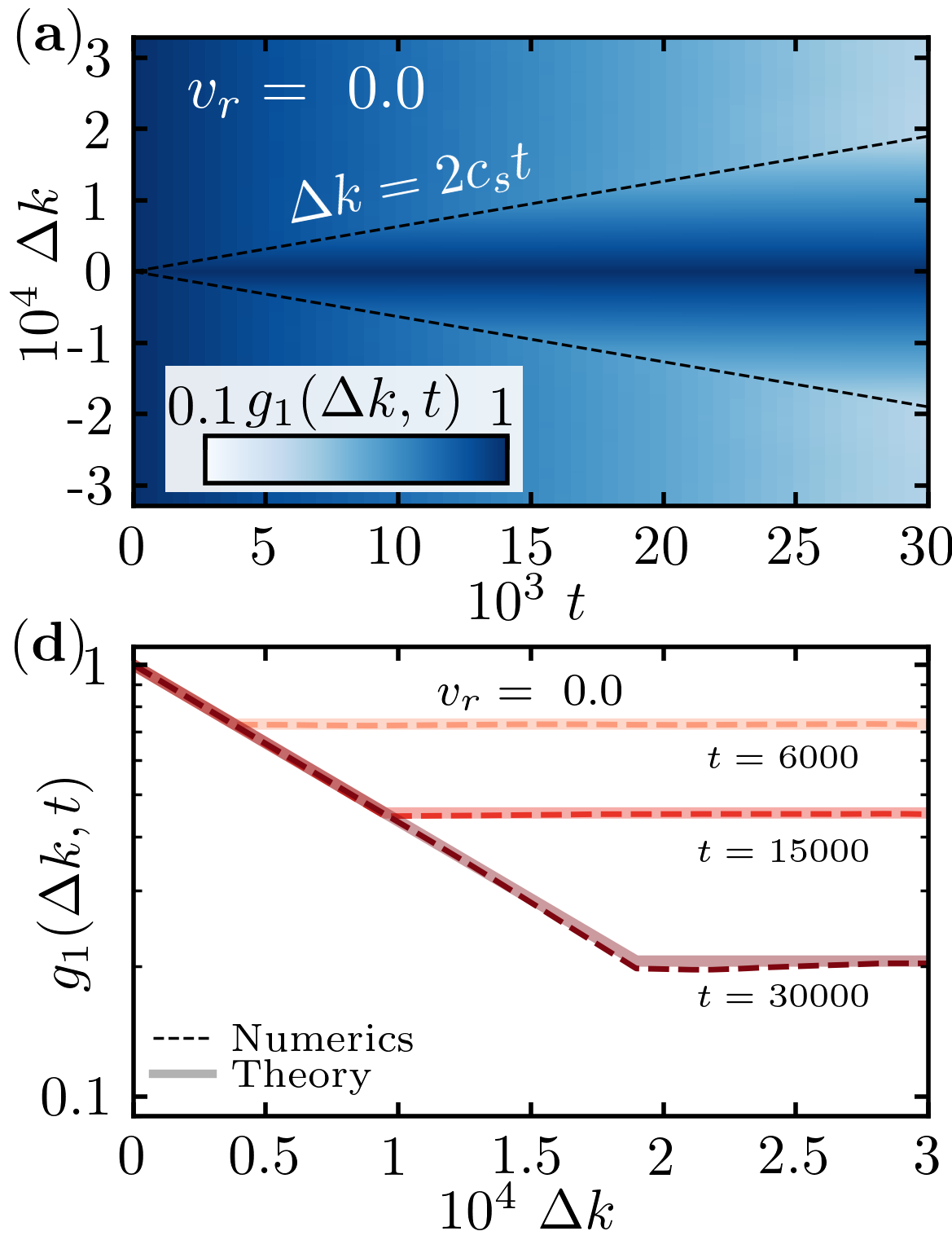}\includegraphics{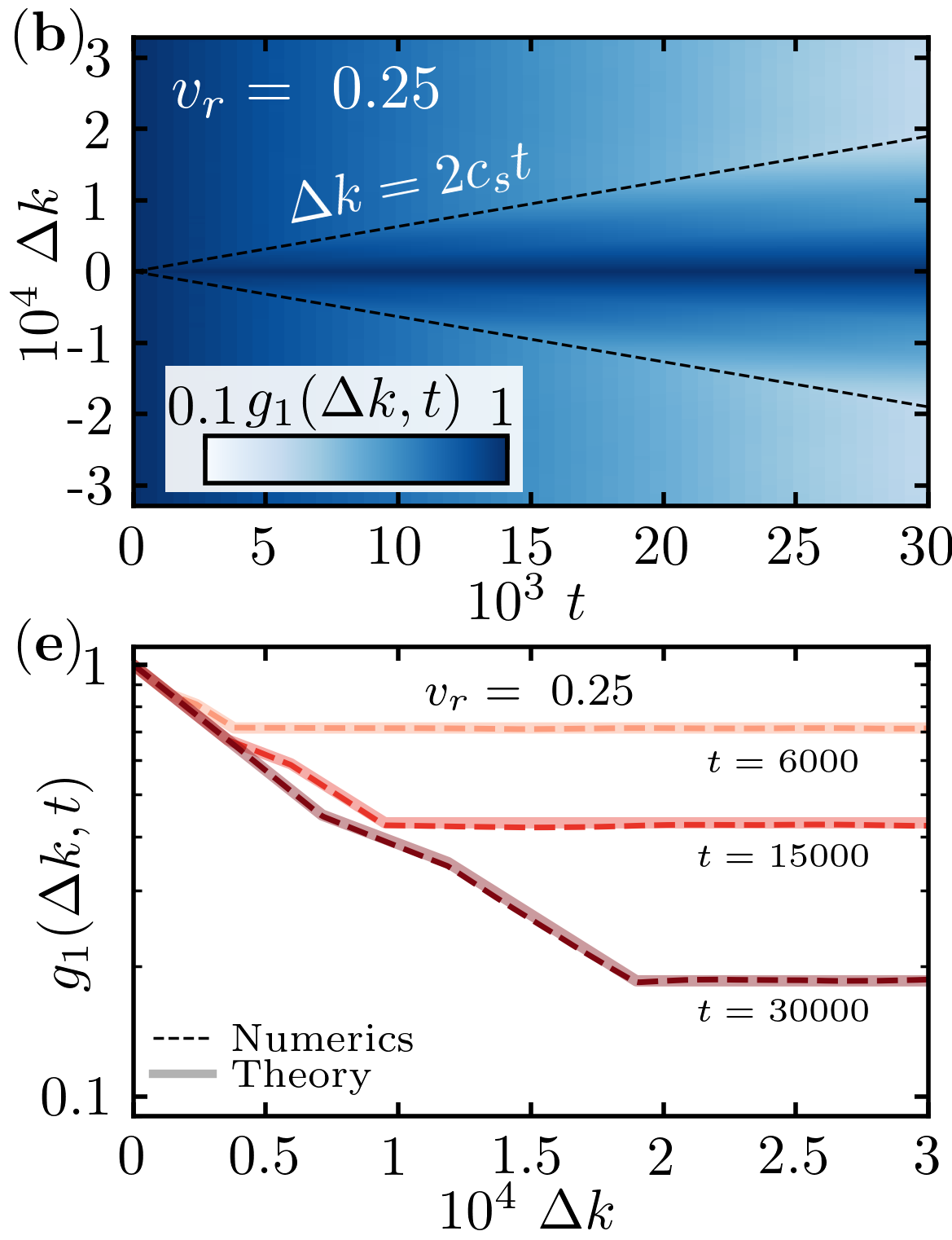}\includegraphics{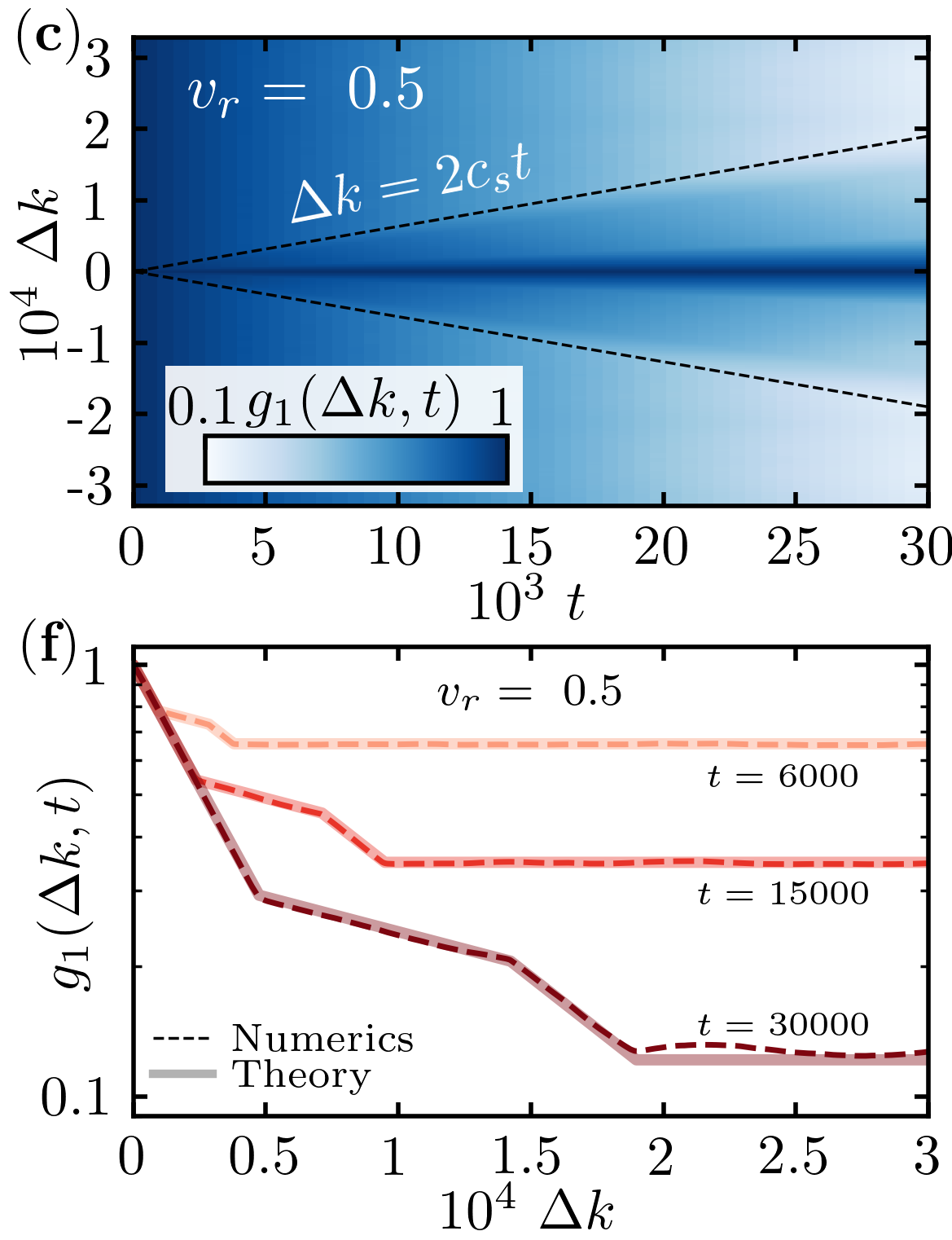}
    \caption{Coherence function $g_1(\Delta k ,t)$, for three different relative velocities $v_r=v/c_s$ and constant speed of sound $c_s\approx 0.31$. Here $t$ is in units of the number of kicks and $\Delta k$ in units of $k_r$. Panels (a-c) are density plots of $g_1$ in the $(t,\Delta k)$ plane  (numerical data), and panels (d-f) are the cuts along the $\Delta k$ axis.
    Corresponding $(K,\phi)$ values are (a,d) $K=0.02$, $\phi=0$, (b,e) $K=0.081$, $\phi=1.32$ and (c,f) $K=0.159$, $\phi=1.44$. All other parameters are fixed: $W=0.02$, $g=5$, $N=65536$ and $n_d=720$ disorder realizations. 
    In panels (a,c), the dashed lines indicate the position of the Lieb-Robinson bound $\Delta k = 2 c_s t$. In panels (d-f), dashed lines are numerical data, while solid thick ones are the theoretical prediction, Eq.~\eqref{eq:g1-integrale-approx-2}, with no adjustable parameter.
  }
    \label{fig:g1_short_time}
\end{figure*}

\subsection{Results}
We show in Fig.~\ref{fig:g1_short_time}
numerical simulations of the coherence function $g_1(\Delta k,t)$, with disorder and interaction parameters chosen such that the NKR lies in the pre-thermal phase: $W/(g\rho_0)\simeq 4\times  10^{-3}$. The panels (a-c) show density plots of the function in the $(t,\Delta k)$ plane for increasing values of $v_r$. At $v_r=0$, the light-cone and the Lieb-Robinson bound $2c_s t$, see Eq.~\eqref{eq:lightcone}, are well visible. At $v_r\ne 0$, on the other hand, the light cones display a more complicated structure. The latter is detailed in panels (d-f), which show cuts of the coherence function along the momentum axis at various times. In these cuts, the analytical prediction~\eqref{eq:g1-integrale-approx-2} is shown on top of the numerical data. The two curves nearly coincide at each time, showing that the agreement between theory and numerics is excellent without any fitting parameter. The cuts, in particular, clearly emphasize the various dynamical regimes of Fig.~\ref{Tk_regimes}.

\section{Long-time thermalization}
\label{Sec:crossover_therm}

\subsection{Crossover to thermal equilibrium and boiling}
\label{subsec:crossover_therm}

In the pre-thermal regime discussed so far, the dynamics is entirely governed by independent phonons. In practice, the system remains in this phase as long as interactions between these phonons are negligible, i.e., at times much smaller than the phonon collision time. Beyond this time scale, the system starts to thermalize. In this section, we discuss the dynamical transition toward the thermal phase in the NKR, and make contact with some of the results  obtained in~\cite{Haldar2022}.

\begin{figure*}[t]
    \centering
    \includegraphics{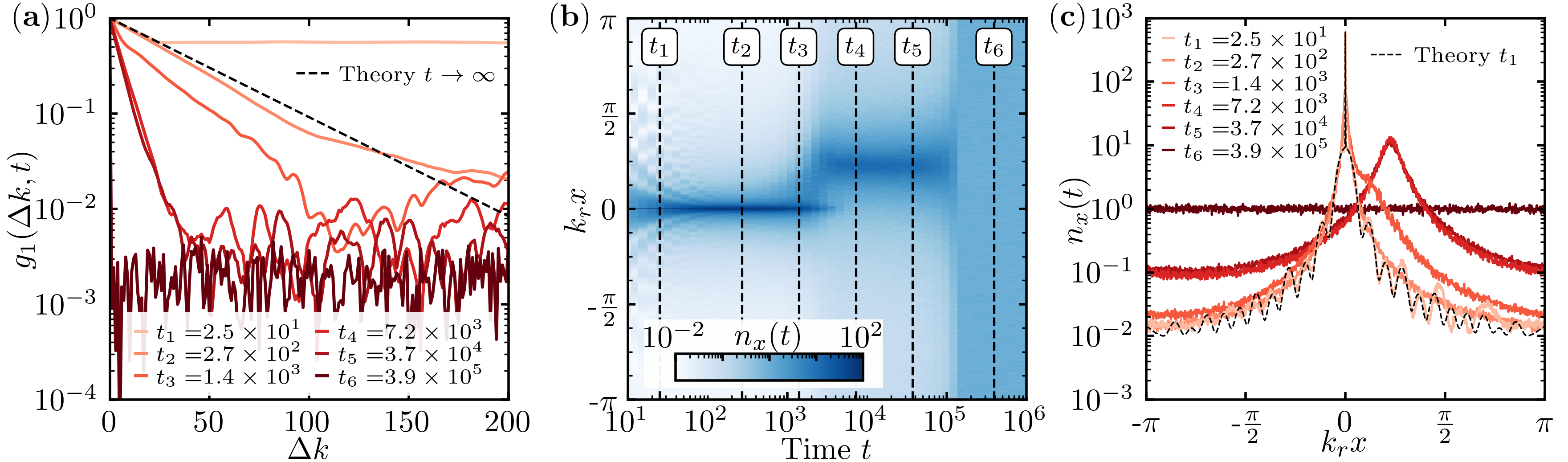}
    \caption{Long-time evolution of the NKR. From $t=0$ onward, the system undergoes three successive dynamical phases. 
    (a) Coherence function $g_1(\Delta k,t)$ as a function of  $\Delta k$, at six successive times $\ t=2.5\times 10^1, 2.7\times 10^2, 1.4\times 10^3, 7.2\times 10^3, 3.7\times 10^4, 3.9\times 10^5$ from top to bottom. Notice, in particular, the time scales much longer than in Fig.~\ref{fig:g1_short_time}. At short time, the function displays the characteristic light-cone spreading of the pre-thermal phase (the dashed line is the  prediction of Bogoliubov theory for $t\to\infty$). At $t\simeq 10^3$, $g_1$ starts decaying faster than the Bogoliubov prediction. At $t\simeq 10^5$, $g_1$ suddenly drops over a single momentum site.
    (b) Density-plot of position distribution $n_x(t)$ as a function of time and position.
    (c) $n_x(t)$ as a function of $x$ for different times, corresponding to vertical cuts of (b) indicated by dashed lines. 
    Parameters are $K=0.10$, $\phi=0.7$, $g=3.0$ (i.e. $v_r\approx 0.13$), $W=0.5$, $N=1024$  and $n_d=180$ disorder realizations.
}
    \label{fig:thermalization-typ}
\end{figure*}

The full dynamical evolution of the NKR in the low-energy limit, i.e., when condition~\eqref{eq:lowenergy} is satisfied, is illustrated by the plots in Fig.~\eqref{fig:thermalization-typ}, where we monitor in time the  coherence function $g_1(\Delta k,t)$ as well as the average position distribution $\smash{n_x(t)=\overline{|\psi(x,t)|^2}}$, which gives a  complementary point of view. 
We recall that these two quantities are simply related through the relation~\eqref{eq:g1_nx}, which explicitly reads
\begin{gather}
\label{eq:n_x_def}
g_1(\Delta k,t) = \int_{-{\pi}/{k_r}}^{{\pi}/{k_r}} \frac{\dd{x}}{2\pi} n_x(t)\e{-i \Delta k x}.
\end{gather}
Note that with respect to the reciprocal system of interacting bosons in a spatially disordered potential discussed in Sec.~\ref{Sec:analogy}, the position distribution $n_x(t)$ here plays the role of the momentum distribution.

In Fig.~\ref{fig:thermalization-typ}(a), we first show the coherence function computed numerically at various times ranging over several decades. For comparison we also show the prediction of Bogoliubov theory at $t\to\infty$ [corresponding to regime~(I) in Eq.~\eqref{eq:g1-integrale-approx-2}]. Whereas the agreement is very good at short time, we observe marked deviations for $t\geq 10^3$. At such long times, $g_1(\Delta k,t)$ starts to decrease faster than the prediction~\eqref{eq:g1-integrale-approx-2}, although the decays appears to remain exponential. This indicates a more significant loss of coherence, which we attribute to the presence of inelastic collisions between phonons. For $t\geq 10^5$, finally, we observe a second change of behavior, where the coherence function abruptly drops from unity to zero over a single momentum site $\Delta k\simeq k_r$.

\begin{figure}
\centering
    \includegraphics{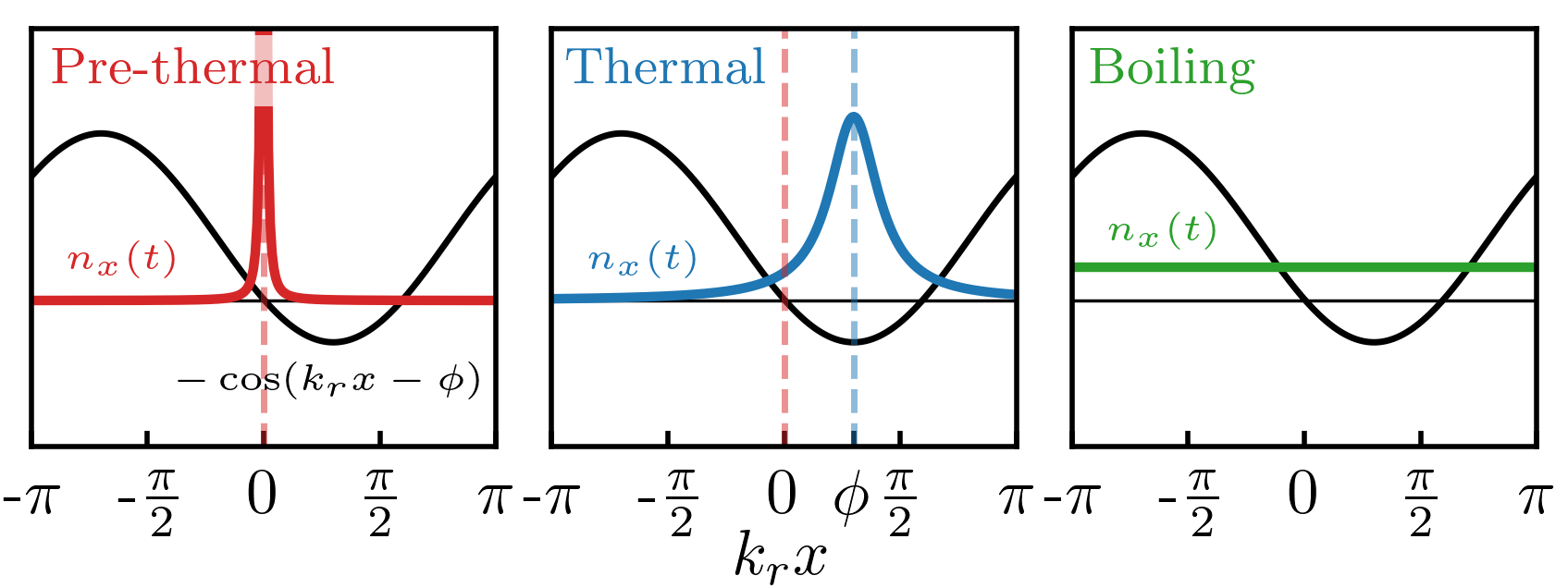}
    \caption{\label{Fig:sketch}
    Sketch showing the position distribution $n_x(t)$ on top of the spatial profile of the kicking potential, $-K\cos(k_rx-\phi)$, in the three dynamical phases (vertical axis is in arbitrary units). In the pre-thermal phase (left panel), the physics is controlled by low-lying excitations corresponding to $n_x(t)$ being centered around $x=0$. In the thermal phase (middle panel), quasi-particles interactions allow the system to populate a broader range of positions, and $n_x(t)$ becomes centered on the potential minimum $x=\phi/k_r$. In the `boiling' phase, finally (right panel), the system heats to infinite temperature and all positions become equally populated. 
    }
\end{figure}

To better understand these results, we show in Fig.~\ref{fig:thermalization-typ}(b) a density plot of the average position distribution $n_x(t)$ in the $(x,t)$ plane, pinpointing the location of the various times considered in Fig.~\ref{fig:thermalization-typ}(a). 
The plot clearly showcases the succession of three dynamical phases as time grows. First, between $t=0$ and $\smash{t\simeq 10^3}$, $n_x(t)$ is rather narrow and centered around $x=0$. This is the pre-thermal phase discussed in the previous sections, where the physics is dominated by low $x-$values. Then, from $t\simeq 10^3$ to $10^5$, the distribution broadens and becomes centered at a nonzero position that turns out to be $x=\phi/k_r$. Finally, beyond $t\simeq 10^5$, $n_x(t)$ uniformly covers the configuration space. The precise spatial profiles of  $n_x(t)$ in these three regimes are shown in Fig.~\ref{fig:thermalization-typ}(c). In the pre-thermal phase (lower red curve at $t=t_1=25$), we also  display the theoretical prediction for $n_x(t)$, calculated using Eqs.~\eqref{eq:n_x_def} and~\eqref{eq:g1-integrale-approx-2}: the distribution is indeed peaked around $x=0$. Its tails also display oscillations stemming from the various interference between phonons at energies $E_x\pm v\hbar x$ [see Eq.~\eqref{eq:g1-integrale}].
At  $t\sim 10^3$, the pre-thermal phase comes to an end, indicating that phonon interactions start to govern the dynamics and that the system thermalizes. Observe, however, that the crossover from the pre-thermal to the thermal phase occurs within a relatively short time window. In the thermal phase, $n_x(t)$ is practically independent of time and acquires  a Lorenztian shape, corresponding to the Rayleigh-Jeans thermal distribution that describes thermal equilibrium in classical-field systems~\cite{Connaughton2005, Cherroret2015, Scoquart2020b}. The fact that the position distribution is centered around $x=\phi/k_r$ in the thermal phase can be elucidated by visualizing the kicking potential of Eq.~\eqref{eq:kicked-ham} on top of $n_x(t)$, as is sketched in Fig.~\ref{Fig:sketch}. In the pre-thermal phase, all the physics is dominated by low-lying excitations, such that the particles remain located near $x=0$ despite the presence of the kicks. When the thermal phase sets in on the contrary,  inelastic quasi-particle collisions occurs and $x\ne 0$ states become accessible. In practice, the particles then spatially occupy the vicinity of the minimum of the kicking potential $-K\cos(k_r x-\phi)$, located at $x=\phi/k_r$, see Fig.~\ref{Fig:sketch}. The thermal phase discussed here was originally observed in~\cite{Haldar2022}. Because a larger ratio $W/(g\rho_0)$ was considered, however, \emph{no} pre-thermal phase involving independent quasi-particles and preceding the thermal regime was observed in that work (although the term `pre-thermal' was  used, somewhat abusively in our opinion).

\begin{figure*}[t]
    \centering
    \includegraphics{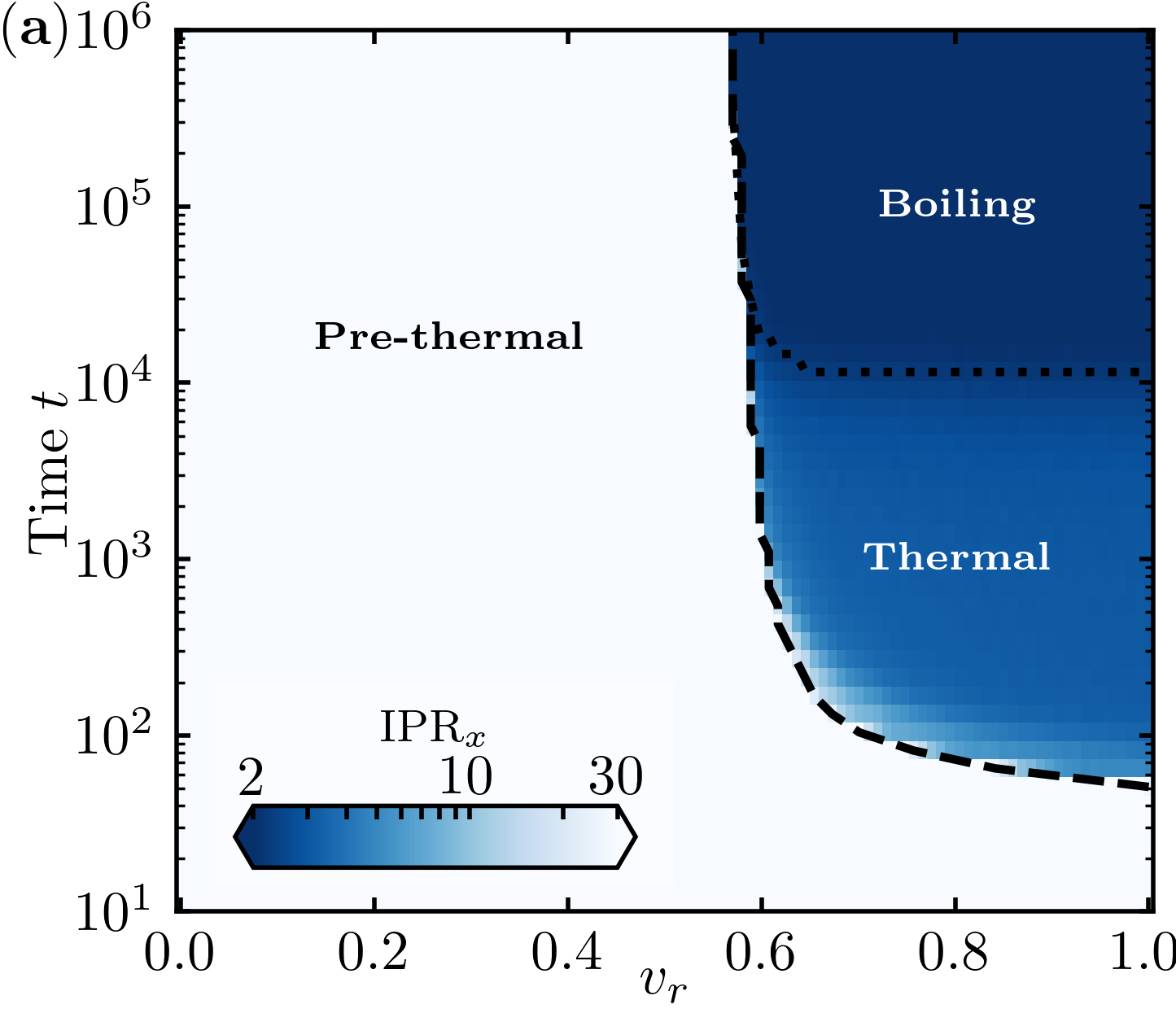}\includegraphics{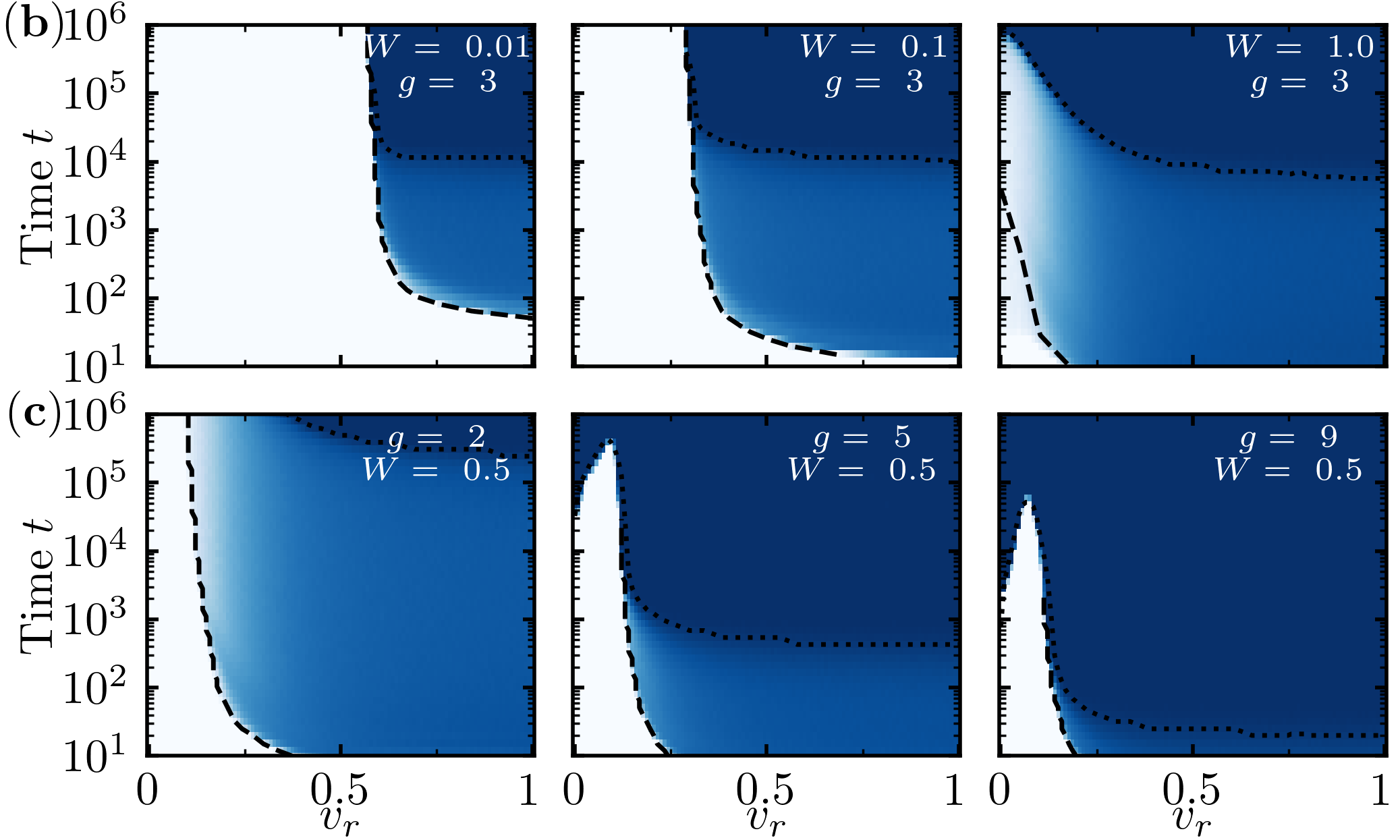}
    \caption{\label{fig:v-crit}Dynamical phase diagram of the NKR vs $v_r=v/c_s$ in the low-energy limit [$W^2/(g\rho_0)^2\ll1$]. In all diagrams,  $N=1024$, $K=0.1$ and we use $n_d=100$ disorder realizations. The ratio $v_r=v/c_s$ is varied via  a change of $\phi$ at constant $K$ (the speed of sound $c_s = \sqrt{K \cos \phi}$ is thus not constant along the $v_r$-axis).
    (a) Phase diagram for $W=0.01$ and $g=3.0$.    In the pre-thermal phase, described in Secs.~\ref{Sec:Pre-thermal dynamics} and ~\ref{Sec:numerics}, the IPR$_x$ is large (of the order of system size). On the contrary, in the boiling phase (thermal phase with infinite temperature), the wave function is fully ergodic and IPR$_x=2$. The cross-overs between the three phases are highlighted by dashed and dotted lines.
     (b) Influence of the disorder energy $W$ at fixed $g=3$ on the phase diagram. (c) Influence of the interaction strength $g$ at fixed $W=0.5$  on the phase diagram.}
\end{figure*}

As seen in Figs.~\ref{fig:thermalization-typ}(b)-(c), finally, the thermal phase is only \emph{metastable} in the NKR. At very long times $t\geq 10^5$ (for the chosen parameters), the distribution becomes flat, indicating that particles become able to move without restriction over the whole configuration space despite the cosine form of the kicking potential, see sketch in Fig.~\ref{Fig:sketch}. This behavior was also pointed out in~\cite{Haldar2022}, where it was referred to as a `boiling'. In the boiling phase, inter-band transitions between quasi-energy states of the kicked rotor are no longer inhibited. The system heats to infinite temperature, featuring a time-independent state  with flat (ergodic) position distribution $n_x(t)$ and, correspondingly, a total absence of coherence with $g_1$ decaying over a single site $k_r$. Here too, the crossover between the thermal and the boiling regimes turns out to be relatively fast.

\subsection{Dynamical phase diagram}

\begin{figure}
    \centering
    \includegraphics[scale=0.9]{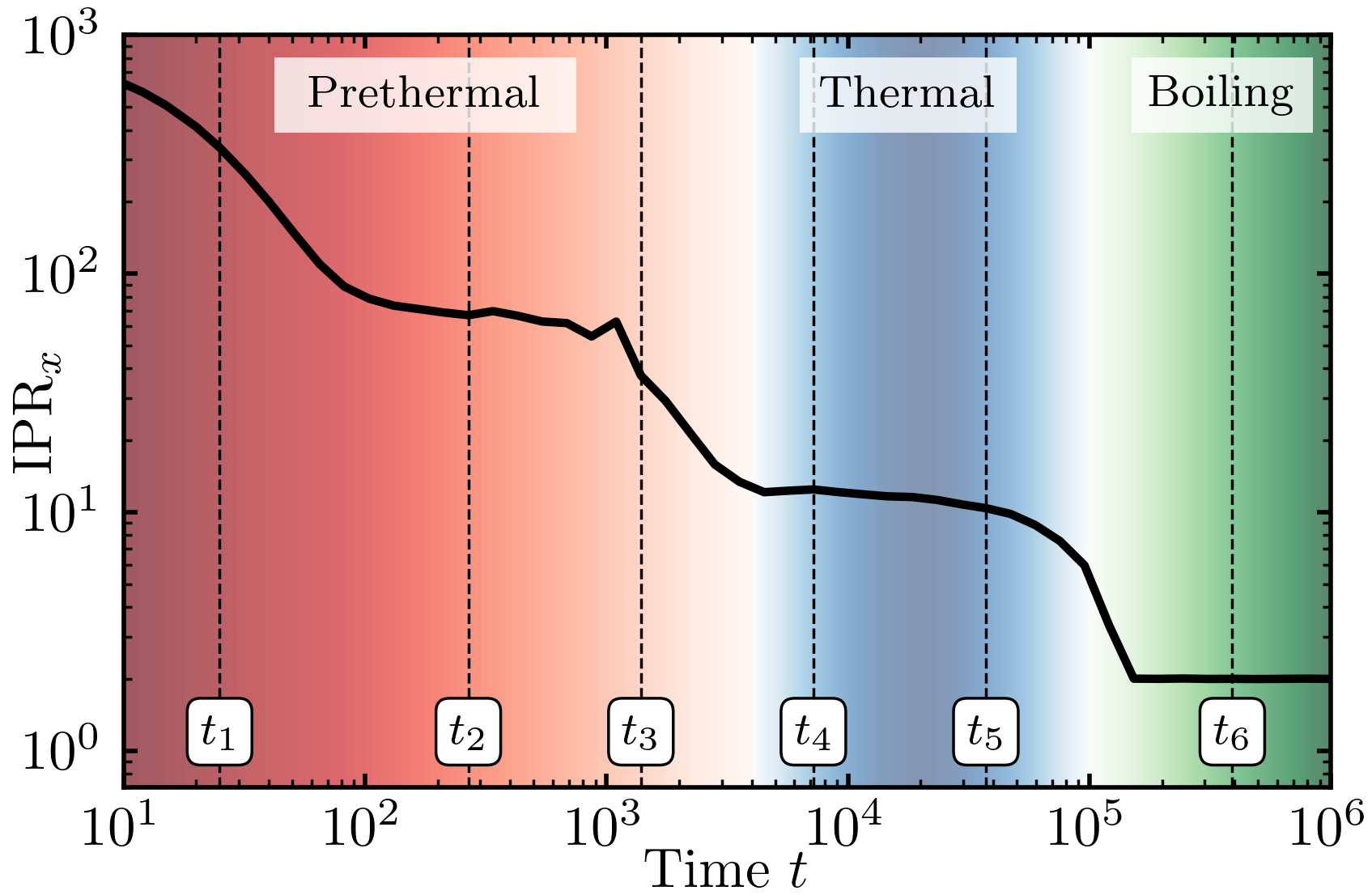}
    \caption{Inverse participation ratio $\text{IPR}_x(t)$ vs. time, when crossing over  the pre-thermal--thermal--boiling phases. At short time, the wave function covers the single state $x=0$ and $\text{IPR}_x(t)\sim N$. In the pre-thermal phase, $\text{IPR}_x(t)$ decreases slowly until it suddenly drops as the system enters  the thermal phase. A second drop at a later time signals the onset of the boiling regime.
   Parameters are the same than in Fig.~\ref{fig:thermalization-typ}: $K=0.1$, $\phi=0.7$, $g=3.0$ (i.e. $v_r\approx 0.13$), $W=1024$ and we use $n_d=180$ disorder realizations.}
    \label{fig:iprx}
\end{figure}
In the previous section, we have described the dynamical crossover from the pre-thermal to the thermal regime, considering a fixed value of the ratio $v_r=v/c_s$, where we recall that $v=K k_r\sin\phi/\hbar$ and $c_s=k_r \sqrt{g\rho_0 K \cos \phi/\hbar^2}$ is the speed of sound. We expect, however, that the dynamics of the NKR in the low-energy limit will be significantly impacted by the value of this ratio. Indeed, remember that, in the hydrodynamic description~\eqref{eq:bog_kr_theta},  $v$ plays the role of a mean velocity in reciprocal space. In the frame of superfluids flowing through a small obstacle, however, it is known from the Landau criterion \cite{Landau1941a, Landau1941b} that the regime $v\sim c_s$ is generally associated with a breakdown of superfluidity \cite{Leggett1999, Pitaevskii2016}. Correspondingly, we have seen in Sec. \ref{Sec:validity} that in the present non-equilibrium context, the validity of the low-energy Bogoliubov description is no longer guaranteed when $v_r\geq 1$ due to the strong increase of density fluctuations. It thus appears natural that a too large $v_r$ value will favor the onset of thermalization. In this section, we quantify precisely this statement, by exploring the full dynamical phase diagram of the NKR when $v_r$ is changed. To this aim, we have carried a large number of temporal propagations of the initial plane-wave state for various values of $v_r$. To change $v_r$, we have chosen to tune the phase $\phi$ at fixed  kick strength $K$. For each $v_r$, we have monitored the dynamical phase using a simple global observable, the inverse participation ratio in position space, IPR$_x$, defined as
\begin{align}
    \text{IPR}_x(t)= \frac{1}{N}\sum_{x} \overline{\vqty{\psi(x,t)}^4 }.
    \label{eq:iprx}
\end{align}
 Use of this quantity is motivated by the idea that the wave function in position space becomes more and more ergodic as the system crosses over from the pre-thermal to the thermal and boiling phases, as already noticed in Sec.~\ref{subsec:crossover_therm}. At $t=0$ and in the early stages of the pre-thermal phase, for instance, $\psi(x,t)$ is concentrated on the single position state $x=0$. Together with the normalization condition~\eqref{eq:normalization_num}, this implies that the inverse participation
$\text{IPR}_x(t)\sim N$ is maximum. In contrast, at very long time in the boiling phase, the wave function uniformly covers the configuration space and becomes a purely random Gaussian variable, such that $\text{IPR}_x(t)=1/N\sum_x 2\overline{|\psi(x,t)|^2}^2=2\overline{|\psi(0,t)|^2}^2=2$ reaches its minimum value. A typical temporal evolution of $\text{IPR}_x(t)$ is shown in Fig.~\ref{fig:iprx}. It exhibits successive drops at the crossovers between the pre-thermal and  thermal phases, and between the thermal and  boiling phases. 

The dynamical phase diagram of the NKR in the $(t,v_r)$ plane is displayed in Fig.~\ref{fig:v-crit}(a). Recall that this diagram is obtained in the low-energy limit where  $W/(g\rho_0)\ll1$. The phase diagram indicates the pre-thermal, thermal and boiling phases. Observe that they are separated by rather sharp crossovers. At very weak $W$ [panel (a)], the pre-thermal phase extends up to extremely long times when $v_r\ll1$. As expected, on the contrary, at $v_r\approx 1$ the system thermalizes at relatively short times, even though we do not observe any particular divergence at $v_r=1$.

Figure~\ref{fig:v-crit}(b) finally shows how the dynamical phase diagram changes as the disorder strength $W$ is increased at fixed interaction strength $g$: at too large $W$, the pre-thermal phase shrinks to zero and only a transition from a thermal to a boiling phase is observed. This is typically the configuration that was considered in~\cite{Haldar2022}. In Fig.~\ref{fig:v-crit}(c), finally, we show the phase diagram at increasing interaction strength $g$ and fixed $W$. It indicates that when the interaction strength is increased too much, the system directly jumps from the pre-thermal to the boiling phase, without intermediate  thermal phase of finite temperature.

\subsection{Critical velocity}

In Fig.~\ref{fig:v-crit}(a), we observe the interesting property that the thermalization time, that is, the time where the pre-thermal dynamics breaks down and leaves room to a thermal equilibrium state, varies extremely rapidly with the fluid velocity around $v_r \simeq 0.6$. In practice, this nearly time-independent value $v_r\equiv v_{r,c} \sim 0.6$ thus acts as an effective critical velocity, below which the system is always pre-thermal, and beyond which it is (almost) always thermal.
\begin{figure}
    \centering
    \includegraphics{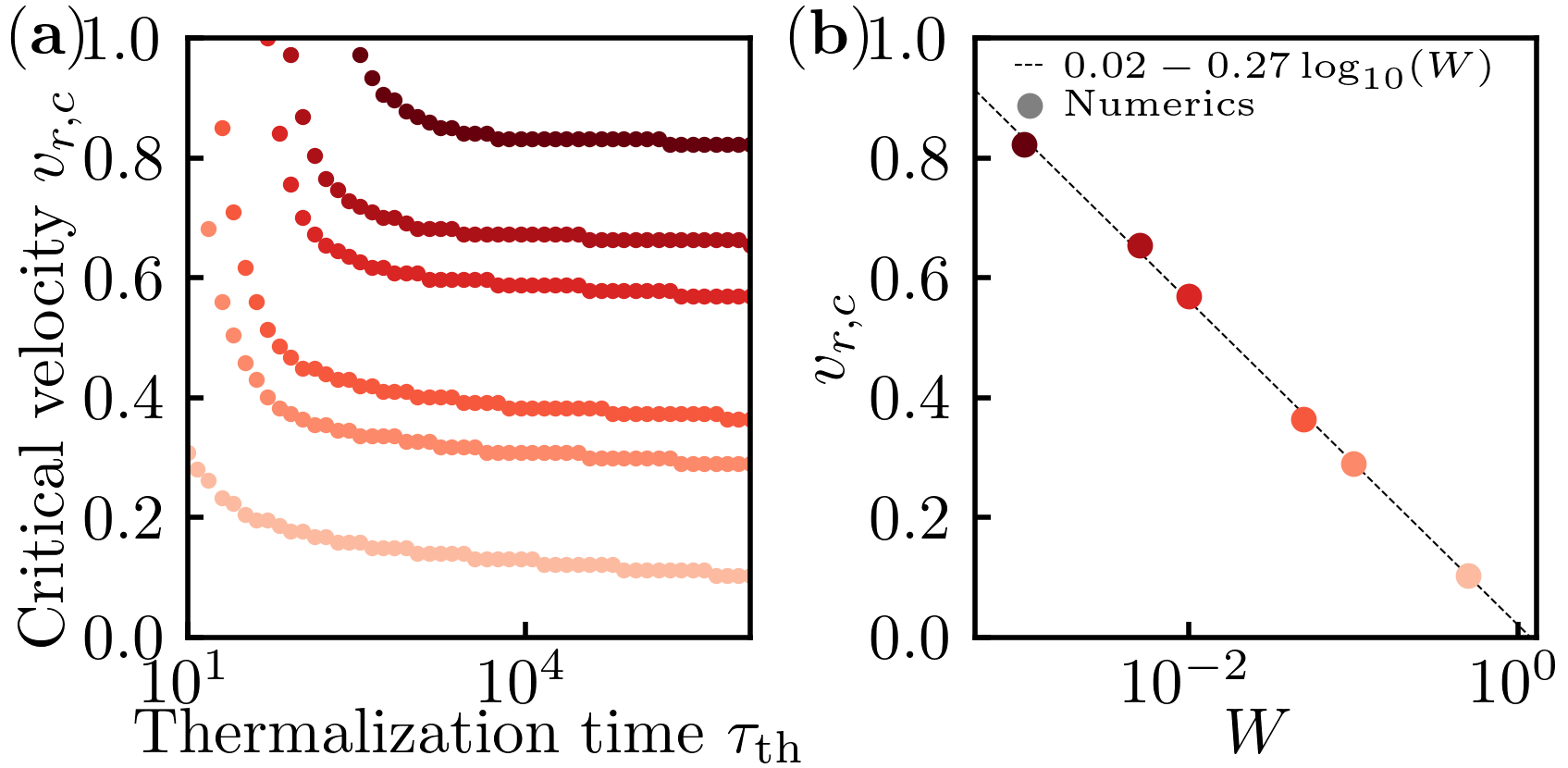}
    \caption{(a) Critical relative velocity $v_{r,c}$ separating the pre-thermal and the thermal/boiling phase as a function of the thermalization time $\tau_\text{th}$ [corresponding to the dashed curve in Fig.~\ref{fig:v-crit}(a)]. The various curves are obtained for different disorder strengths $W=0.5,0.1,0.05,0.01,0.005,0.001$ from top to bottom.
    Other parameters are $K=0.1$, $g=3.0$, $N=1024$ and we have used $n_d=100$ disorder realizations.
    (b) Critical velocity at the longest thermalization time computed, $\tau_\text{th}=10^6$, vs. disorder strength. Symbols are numerical data and the dotted line is a phenomenological logarithmic fit.
}    \label{fig:1st_thermalization_time}
\end{figure}
This critical velocity line is reported in Fig.~\ref{fig:1st_thermalization_time}(a) for various disorder strengths $W$, and in Fig.~\ref{fig:1st_thermalization_time}(b) we show its nearly constant value as a function of $W$. In agreement with the phase diagrams in Figs.~\ref{fig:v-crit}(b), $v_{r,c}$ decreases to zero as $W$ is increased, with a decay that appears to be close to logarithmic.
\section{Conclusion}
\label{Sec:conclusion}

In this article, we have developed a low-energy hydrodynamic theory of the nonlinear kicked rotor with cubic repulsive interactions, and we have used it to describe the quench dynamics of a plane-wave state and the ensuing non-equilibrium pre-thermal regime. We have shown, in particular, that this system is the reciprocal version (in momentum space) of a weakly interacting Bose gas of finite mean velocity in the presence of a spatially disordered potential, the mean velocity being controllable via the phase of the periodic modulation of the kicks. The hydrodynamic approach is valid provided the random kinetic energy is smaller than the interaction energy and the effective mean gas velocity is smaller than the speed of sound. In this regime, we have found an excellent agreement with exact numerical simulations. Finally, we have explored the dynamical crossover from the pre-thermal to the thermal phase that occurs when the evolution time becomes longer than the quasi-particle collision time, and have described how these phases are impacted by the interactions, the fluctuations of the kinetic phases and the effective mean velocity.

\acknowledgments

MM acknowledges Calcul en Midi-Pyrénées (CALMIP), France, for access to its supercomputer. NC and PEL thank Gabriel Lemari\'e, Cl\'ement Duval, and Mathias Albert for discussions. This work has benefited from the financial support of Agence Nationale de la Recherche (ANR), France, under Grants  Nos.~ANR-19-CE30-0028-01 CONFOCAL for NC, ANR-18-CE30-0017 MANYLOK for DD and MM, and  ANR-21-CE47-0009 Quantum-SOPHA for PEL.

\appendix

\section{Conventions}

\label{App:Conventions}

To help the reader throughout the manuscript, we recall here the conventions we use and their numerical implementations.

In Hamiltonian~\eqref{eq:kicked-ham}, position $x$ is defined on the circle $x \in [-\pi/k_r,\pi/k_r]$ with periodic boundary conditions. Wave vector $k$ then takes discrete values $k_l = lk_r $ ($l \in \mathbb{Z}$).

In simulations, we set $\hbar = k_r = T =1$ and work with a finite system size $N$ for the momentum grid, with periodic boundary conditions. The wave numbers $k$ then take integer values  $k =  -N/2+1,\dots 0, \dots N/2$ (for $N$ even). 
Consequently, position $x$ also takes discrete values $x_n=\pm {\pi}/{N}, \pm {3 \pi}/{N}, \dots \pm {(N-1) \pi}/{N}$.

Conventions and numerical implementation for Fourier transform and wave function normalization are given in Table~\ref{table:convention}. 

\begin{table}[!h]
\[\arraycolsep=1.5 pt\def\arraystretch{1.8}
    \begin{array}{c|c}
    \text{Definition} & \text{Numerical implementation} \\
    \hline
    x \in [-\frac{\pi}{k_r},\frac{\pi}{k_r}] & x_n= \pm \frac{\pi}{N}, \dots \pm \frac{(N-1) \pi}{N} \\
    k_l = l k_r ~|~ l \in \mathbb{Z}  & k = - \frac{N}{2}+1,\dots 0, \dots, \frac{N}{2} \\
    \hline
         \int_{-{\pi}/{k_r}}^{{\pi}/{k_r}} \frac{\dd{x}}{2\pi} \vqty*{\tilde{\psi}(x)}^2 =1 &  \frac{1}{N} \sum_{x_n}   \vqty*{\tilde{\psi}(x_n)}^2 =1  \\
         k_r \sum_{k_l=lk_r }  \vqty{\psi(k_l)}^2 =1&  \sum_k \vqty{\psi(k)}^2 =1 \\
         \hline
         \tilde{\psi}(x) = k_r \sum_{k_l=lk_r} \psi(k_l) \e{ik_l x_n} & \tilde{\psi}(x_n)= \sum_k \psi(k) \e{ik x_n} \\
         \psi(k_l) = \int_{-{\pi}/{k_r}}^{{\pi}/{k_r}} \frac{\dd{x}}{2\pi}  \tilde{\psi}(x) \e{-ik_l x} & \psi(k)= \frac{1}{N} \sum_{x_n} \tilde{\psi}(x_n)  \e{-ik x_n} \\
    \end{array}\]
    \caption{\label{table:convention}Definitions used in the manuscript and their numerical implementations.}
\end{table}

\bibliography{biblio.bib}

\begin{thebibliography}{70}%
\makeatletter
\providecommand \@ifxundefined [1]{%
 \@ifx{#1\undefined}
}%
\providecommand \@ifnum [1]{%
 \ifnum #1\expandafter \@firstoftwo
 \else \expandafter \@secondoftwo
 \fi
}%
\providecommand \@ifx [1]{%
 \ifx #1\expandafter \@firstoftwo
 \else \expandafter \@secondoftwo
 \fi
}%
\providecommand \natexlab [1]{#1}%
\providecommand \enquote  [1]{``#1''}%
\providecommand \bibnamefont  [1]{#1}%
\providecommand \bibfnamefont [1]{#1}%
\providecommand \citenamefont [1]{#1}%
\providecommand \href@noop [0]{\@secondoftwo}%
\providecommand \href [0]{\begingroup \@sanitize@url \@href}%
\providecommand \@href[1]{\@@startlink{#1}\@@href}%
\providecommand \@@href[1]{\endgroup#1\@@endlink}%
\providecommand \@sanitize@url [0]{\catcode `\\12\catcode `\$12\catcode
  `\&12\catcode `\#12\catcode `\^12\catcode `\_12\catcode `\%12\relax}%
\providecommand \@@startlink[1]{}%
\providecommand \@@endlink[0]{}%
\providecommand \url  [0]{\begingroup\@sanitize@url \@url }%
\providecommand \@url [1]{\endgroup\@href {#1}{\urlprefix }}%
\providecommand \urlprefix  [0]{URL }%
\providecommand \Eprint [0]{\href }%
\providecommand \doibase [0]{http://dx.doi.org/}%
\providecommand \selectlanguage [0]{\@gobble}%
\providecommand \bibinfo  [0]{\@secondoftwo}%
\providecommand \bibfield  [0]{\@secondoftwo}%
\providecommand \translation [1]{[#1]}%
\providecommand \BibitemOpen [0]{}%
\providecommand \bibitemStop [0]{}%
\providecommand \bibitemNoStop [0]{.\EOS\space}%
\providecommand \EOS [0]{\spacefactor3000\relax}%
\providecommand \BibitemShut  [1]{\csname bibitem#1\endcsname}%
\let\auto@bib@innerbib\@empty
\bibitem [{\citenamefont {Polkovnikov}\ \emph {et~al.}(2011)\citenamefont
  {Polkovnikov}, \citenamefont {Sengupta}, \citenamefont {Silva},\ and\
  \citenamefont {Vengalattore}}]{Polkovnikov2011}%
  \BibitemOpen
  \bibfield  {author} {\bibinfo {author} {\bibfnamefont {A.}~\bibnamefont
  {Polkovnikov}}, \bibinfo {author} {\bibfnamefont {K.}~\bibnamefont
  {Sengupta}}, \bibinfo {author} {\bibfnamefont {A.}~\bibnamefont {Silva}}, \
  and\ \bibinfo {author} {\bibfnamefont {M.}~\bibnamefont {Vengalattore}},\
  }\href {https://doi.org/10.1103/RevModPhys.83.863} {\bibfield  {journal}
  {\bibinfo  {journal} {Rev. Mod. Phys.}\ }\textbf {\bibinfo {volume} {83}},\
  \bibinfo {pages} {863} (\bibinfo {year} {2011})}\BibitemShut {NoStop}%
\bibitem [{\citenamefont {Gogolin}\ and\ \citenamefont
  {Eisert}(2016)}]{Gogolin2016}%
  \BibitemOpen
  \bibfield  {author} {\bibinfo {author} {\bibfnamefont {C.}~\bibnamefont
  {Gogolin}}\ and\ \bibinfo {author} {\bibfnamefont {J.}~\bibnamefont
  {Eisert}},\ }\href {https://doi.org/10.1088/0034-4885/79/5/056001} {\bibfield
   {journal} {\bibinfo  {journal} {Rep. Prog. Phys.}\ }\textbf {\bibinfo
  {volume} {79}},\ \bibinfo {pages} {056001} (\bibinfo {year}
  {2016})}\BibitemShut {NoStop}%
\bibitem [{\citenamefont {D'Alessio}\ \emph {et~al.}(2016)\citenamefont
  {D'Alessio}, \citenamefont {Kafri}, \citenamefont {Polkovnikov},\ and\
  \citenamefont {Rigol}}]{Alessio2016}%
  \BibitemOpen
  \bibfield  {author} {\bibinfo {author} {\bibfnamefont {L.}~\bibnamefont
  {D'Alessio}}, \bibinfo {author} {\bibfnamefont {Y.}~\bibnamefont {Kafri}},
  \bibinfo {author} {\bibfnamefont {A.}~\bibnamefont {Polkovnikov}}, \ and\
  \bibinfo {author} {\bibfnamefont {M.}~\bibnamefont {Rigol}},\ }\href
  {https://doi.org/10.1080/00018732.2016.1198134} {\bibfield  {journal}
  {\bibinfo  {journal} {Adv. Phys.}\ }\textbf {\bibinfo {volume} {65}},\
  \bibinfo {pages} {239} (\bibinfo {year} {2016})}\BibitemShut {NoStop}%
\bibitem [{\citenamefont {Deutsch}(2018)}]{Deutsch2018}%
  \BibitemOpen
  \bibfield  {author} {\bibinfo {author} {\bibfnamefont {J.~M.}\ \bibnamefont
  {Deutsch}},\ }\href {https://doi.org/10.1088/1361-6633/aac9f1} {\bibfield
  {journal} {\bibinfo  {journal} {Rep. Prog. Phys.}\ }\textbf {\bibinfo
  {volume} {81}},\ \bibinfo {pages} {082001} (\bibinfo {year}
  {2018})}\BibitemShut {NoStop}%
\bibitem [{\citenamefont {Berges}\ \emph {et~al.}(2004)\citenamefont {Berges},
  \citenamefont {Bors{\'{a}}nyi},\ and\ \citenamefont
  {Wetterich}}]{Berges2004}%
  \BibitemOpen
  \bibfield  {author} {\bibinfo {author} {\bibfnamefont {J.}~\bibnamefont
  {Berges}}, \bibinfo {author} {\bibfnamefont {S.}~\bibnamefont
  {Bors{\'{a}}nyi}}, \ and\ \bibinfo {author} {\bibfnamefont {C.}~\bibnamefont
  {Wetterich}},\ }\href {https://doi.org/10.1103%2Fphysrevlett.93.142002}
  {\bibfield  {journal} {\bibinfo  {journal} {Phys. Rev. Lett.}\ }\textbf
  {\bibinfo {volume} {93}},\ \bibinfo {pages} {142002} (\bibinfo {year}
  {2004})}\BibitemShut {NoStop}%
\bibitem [{\citenamefont {Kitagawa}\ \emph {et~al.}(2011)\citenamefont
  {Kitagawa}, \citenamefont {Imambekov}, \citenamefont {Schmiedmayer},\ and\
  \citenamefont {Demler}}]{Kitagawa2011}%
  \BibitemOpen
  \bibfield  {author} {\bibinfo {author} {\bibfnamefont {T.}~\bibnamefont
  {Kitagawa}}, \bibinfo {author} {\bibfnamefont {A.}~\bibnamefont {Imambekov}},
  \bibinfo {author} {\bibfnamefont {J.}~\bibnamefont {Schmiedmayer}}, \ and\
  \bibinfo {author} {\bibfnamefont {E.}~\bibnamefont {Demler}},\ }\href
  {https://doi.org/10.1088/1367-2630/13/7/073018} {\bibfield  {journal}
  {\bibinfo  {journal} {New J. Phys.}\ }\textbf {\bibinfo {volume} {13}},\
  \bibinfo {pages} {073018} (\bibinfo {year} {2011})}\BibitemShut {NoStop}%
\bibitem [{\citenamefont {Kollar}\ \emph {et~al.}(2011)\citenamefont {Kollar},
  \citenamefont {Wolf},\ and\ \citenamefont {Eckstein}}]{Kollar2011}%
  \BibitemOpen
  \bibfield  {author} {\bibinfo {author} {\bibfnamefont {M.}~\bibnamefont
  {Kollar}}, \bibinfo {author} {\bibfnamefont {F.~A.}\ \bibnamefont {Wolf}}, \
  and\ \bibinfo {author} {\bibfnamefont {M.}~\bibnamefont {Eckstein}},\ }\href
  {https://doi.org/10.1103%2Fphysrevb.84.054304} {\bibfield  {journal}
  {\bibinfo  {journal} {Phys. Rev. B}\ }\textbf {\bibinfo {volume} {84}},\
  \bibinfo {pages} {054304} (\bibinfo {year} {2011})}\BibitemShut {NoStop}%
\bibitem [{\citenamefont {Buchhold}\ \emph {et~al.}(2016)\citenamefont
  {Buchhold}, \citenamefont {Heyl},\ and\ \citenamefont
  {Diehl}}]{Buchhold2016}%
  \BibitemOpen
  \bibfield  {author} {\bibinfo {author} {\bibfnamefont {M.}~\bibnamefont
  {Buchhold}}, \bibinfo {author} {\bibfnamefont {M.}~\bibnamefont {Heyl}}, \
  and\ \bibinfo {author} {\bibfnamefont {S.}~\bibnamefont {Diehl}},\ }\href
  {https://doi.org/10.1103/PhysRevA.94.013601} {\bibfield  {journal} {\bibinfo
  {journal} {Phys. Rev. A}\ }\textbf {\bibinfo {volume} {94}},\ \bibinfo
  {pages} {013601} (\bibinfo {year} {2016})}\BibitemShut {NoStop}%
\bibitem [{\citenamefont {Larr{\'{e}}}\ \emph {et~al.}(2018)\citenamefont
  {Larr{\'{e}}}, \citenamefont {Delande},\ and\ \citenamefont
  {Cherroret}}]{Larre2018}%
  \BibitemOpen
  \bibfield  {author} {\bibinfo {author} {\bibfnamefont {P.-{\'{E}}.}\
  \bibnamefont {Larr{\'{e}}}}, \bibinfo {author} {\bibfnamefont
  {D.}~\bibnamefont {Delande}}, \ and\ \bibinfo {author} {\bibfnamefont
  {N.}~\bibnamefont {Cherroret}},\ }\href
  {https://doi.org/10.1103%2Fphysreva.97.043805} {\bibfield  {journal}
  {\bibinfo  {journal} {Phys. Rev. A}\ }\textbf {\bibinfo {volume} {97}},\
  \bibinfo {pages} {043805} (\bibinfo {year} {2018})}\BibitemShut {NoStop}%
\bibitem [{\citenamefont {Mori}\ \emph {et~al.}(2018)\citenamefont {Mori},
  \citenamefont {Ikeda}, \citenamefont {Kaminishi},\ and\ \citenamefont
  {Ueda}}]{Mori2018}%
  \BibitemOpen
  \bibfield  {author} {\bibinfo {author} {\bibfnamefont {T.}~\bibnamefont
  {Mori}}, \bibinfo {author} {\bibfnamefont {T.~N.}\ \bibnamefont {Ikeda}},
  \bibinfo {author} {\bibfnamefont {E.}~\bibnamefont {Kaminishi}}, \ and\
  \bibinfo {author} {\bibfnamefont {M.}~\bibnamefont {Ueda}},\ }\href
  {https://doi.org/10.1088%2F1361-6455%2Faabcdf} {\bibfield  {journal}
  {\bibinfo  {journal} {J. Phys. B: At. Mol. Opt. Phys.}\ }\textbf {\bibinfo
  {volume} {51}},\ \bibinfo {pages} {112001} (\bibinfo {year}
  {2018})}\BibitemShut {NoStop}%
\bibitem [{\citenamefont {Martone}\ \emph {et~al.}(2018)\citenamefont
  {Martone}, \citenamefont {Larr{\'e}}, \citenamefont {Fabbri},\ and\
  \citenamefont {Pavloff}}]{Martone2018}%
  \BibitemOpen
  \bibfield  {author} {\bibinfo {author} {\bibfnamefont {G.~I.}\ \bibnamefont
  {Martone}}, \bibinfo {author} {\bibfnamefont {P.-{\'E}.}\ \bibnamefont
  {Larr{\'e}}}, \bibinfo {author} {\bibfnamefont {A.}~\bibnamefont {Fabbri}}, \
  and\ \bibinfo {author} {\bibfnamefont {N.}~\bibnamefont {Pavloff}},\ }\href
  {https://doi.org/10.1103/PhysRevA.98.063617} {\bibfield  {journal} {\bibinfo
  {journal} {Phys. Rev. A}\ }\textbf {\bibinfo {volume} {98}},\ \bibinfo
  {pages} {063617} (\bibinfo {year} {2018})}\BibitemShut {NoStop}%
\bibitem [{\citenamefont {Mallayya}\ \emph {et~al.}(2019)\citenamefont
  {Mallayya}, \citenamefont {Rigol},\ and\ \citenamefont
  {{De~Roeck}}}]{Mallayya2019}%
  \BibitemOpen
  \bibfield  {author} {\bibinfo {author} {\bibfnamefont {K.}~\bibnamefont
  {Mallayya}}, \bibinfo {author} {\bibfnamefont {M.}~\bibnamefont {Rigol}}, \
  and\ \bibinfo {author} {\bibfnamefont {W.}~\bibnamefont {{De~Roeck}}},\
  }\href {https://doi.org/10.1103%2Fphysrevx.9.021027} {\bibfield  {journal}
  {\bibinfo  {journal} {Phys. Rev. X}\ }\textbf {\bibinfo {volume} {9}},\
  \bibinfo {pages} {021027} (\bibinfo {year} {2019})}\BibitemShut {NoStop}%
\bibitem [{\citenamefont {Gring}\ \emph {et~al.}(2012)\citenamefont {Gring},
  \citenamefont {Kuhnert}, \citenamefont {Langen}, \citenamefont {Kitagawa},
  \citenamefont {Rauer}, \citenamefont {Schreitl}, \citenamefont {Mazets},
  \citenamefont {Smith}, \citenamefont {Demler},\ and\ \citenamefont
  {Schmiedmayer}}]{Gring2012}%
  \BibitemOpen
  \bibfield  {author} {\bibinfo {author} {\bibfnamefont {M.}~\bibnamefont
  {Gring}}, \bibinfo {author} {\bibfnamefont {M.}~\bibnamefont {Kuhnert}},
  \bibinfo {author} {\bibfnamefont {T.}~\bibnamefont {Langen}}, \bibinfo
  {author} {\bibfnamefont {T.}~\bibnamefont {Kitagawa}}, \bibinfo {author}
  {\bibfnamefont {B.}~\bibnamefont {Rauer}}, \bibinfo {author} {\bibfnamefont
  {M.}~\bibnamefont {Schreitl}}, \bibinfo {author} {\bibfnamefont
  {I.}~\bibnamefont {Mazets}}, \bibinfo {author} {\bibfnamefont {D.~A.}\
  \bibnamefont {Smith}}, \bibinfo {author} {\bibfnamefont {E.}~\bibnamefont
  {Demler}}, \ and\ \bibinfo {author} {\bibfnamefont {J.}~\bibnamefont
  {Schmiedmayer}},\ }\href {https://doi.org/10.1126%2Fscience.1224953}
  {\bibfield  {journal} {\bibinfo  {journal} {Science}\ }\textbf {\bibinfo
  {volume} {337}},\ \bibinfo {pages} {1318} (\bibinfo {year}
  {2012})}\BibitemShut {NoStop}%
\bibitem [{\citenamefont {Langen}\ \emph {et~al.}(2013)\citenamefont {Langen},
  \citenamefont {Geiger}, \citenamefont {Kuhnert}, \citenamefont {Rauer},\ and\
  \citenamefont {Schmiedmayer}}]{Langen2013}%
  \BibitemOpen
  \bibfield  {author} {\bibinfo {author} {\bibfnamefont {T.}~\bibnamefont
  {Langen}}, \bibinfo {author} {\bibfnamefont {R.}~\bibnamefont {Geiger}},
  \bibinfo {author} {\bibfnamefont {M.}~\bibnamefont {Kuhnert}}, \bibinfo
  {author} {\bibfnamefont {B.}~\bibnamefont {Rauer}}, \ and\ \bibinfo {author}
  {\bibfnamefont {J.}~\bibnamefont {Schmiedmayer}},\ }\href
  {https://doi.org/10.1038%2Fnphys2739} {\bibfield  {journal} {\bibinfo
  {journal} {Nat. Phys.}\ }\textbf {\bibinfo {volume} {9}},\ \bibinfo {pages}
  {640} (\bibinfo {year} {2013})}\BibitemShut {NoStop}%
\bibitem [{\citenamefont {Langen}\ \emph {et~al.}(2015)\citenamefont {Langen},
  \citenamefont {Erne}, \citenamefont {Geiger}, \citenamefont {Rauer},
  \citenamefont {Schweigler}, \citenamefont {Kuhnert}, \citenamefont
  {Rohringer}, \citenamefont {Mazets}, \citenamefont {Gasenzer},\ and\
  \citenamefont {Schmiedmayer}}]{Langen2015}%
  \BibitemOpen
  \bibfield  {author} {\bibinfo {author} {\bibfnamefont {T.}~\bibnamefont
  {Langen}}, \bibinfo {author} {\bibfnamefont {S.}~\bibnamefont {Erne}},
  \bibinfo {author} {\bibfnamefont {R.}~\bibnamefont {Geiger}}, \bibinfo
  {author} {\bibfnamefont {B.}~\bibnamefont {Rauer}}, \bibinfo {author}
  {\bibfnamefont {T.}~\bibnamefont {Schweigler}}, \bibinfo {author}
  {\bibfnamefont {M.}~\bibnamefont {Kuhnert}}, \bibinfo {author} {\bibfnamefont
  {W.}~\bibnamefont {Rohringer}}, \bibinfo {author} {\bibfnamefont {I.~E.}\
  \bibnamefont {Mazets}}, \bibinfo {author} {\bibfnamefont {T.}~\bibnamefont
  {Gasenzer}}, \ and\ \bibinfo {author} {\bibfnamefont {J.}~\bibnamefont
  {Schmiedmayer}},\ }\href {https://doi.org/10.1126%2Fscience.1257026}
  {\bibfield  {journal} {\bibinfo  {journal} {Science}\ }\textbf {\bibinfo
  {volume} {348}},\ \bibinfo {pages} {207} (\bibinfo {year}
  {2015})}\BibitemShut {NoStop}%
\bibitem [{\citenamefont {Abuzarli}\ \emph {et~al.}(2022)\citenamefont
  {Abuzarli}, \citenamefont {Cherroret}, \citenamefont {Bienaim\'e},\ and\
  \citenamefont {Glorieux}}]{Abuzarli2022}%
  \BibitemOpen
  \bibfield  {author} {\bibinfo {author} {\bibfnamefont {M.}~\bibnamefont
  {Abuzarli}}, \bibinfo {author} {\bibfnamefont {N.}~\bibnamefont {Cherroret}},
  \bibinfo {author} {\bibfnamefont {T.}~\bibnamefont {Bienaim\'e}}, \ and\
  \bibinfo {author} {\bibfnamefont {Q.}~\bibnamefont {Glorieux}},\ }\href
  {\doibase 10.1103/PhysRevLett.129.100602} {\bibfield  {journal} {\bibinfo
  {journal} {Phys. Rev. Lett.}\ }\textbf {\bibinfo {volume} {129}},\ \bibinfo
  {pages} {100602} (\bibinfo {year} {2022})}\BibitemShut {NoStop}%
\bibitem [{\citenamefont {Reitter}\ \emph {et~al.}(2017)\citenamefont
  {Reitter}, \citenamefont {Näger}, \citenamefont {Wintersperger},
  \citenamefont {Sträter}, \citenamefont {Bloch}, \citenamefont {Eckardt},\
  and\ \citenamefont {Schneider}}]{Reitter2017}%
  \BibitemOpen
  \bibfield  {author} {\bibinfo {author} {\bibfnamefont {M.}~\bibnamefont
  {Reitter}}, \bibinfo {author} {\bibfnamefont {J.}~\bibnamefont {Näger}},
  \bibinfo {author} {\bibfnamefont {K.}~\bibnamefont {Wintersperger}}, \bibinfo
  {author} {\bibfnamefont {C.}~\bibnamefont {Sträter}}, \bibinfo {author}
  {\bibfnamefont {I.}~\bibnamefont {Bloch}}, \bibinfo {author} {\bibfnamefont
  {A.}~\bibnamefont {Eckardt}}, \ and\ \bibinfo {author} {\bibfnamefont
  {U.}~\bibnamefont {Schneider}},\ }\href
  {https://doi.org/10.1103%2Fphysrevlett.119.200402} {\bibfield  {journal}
  {\bibinfo  {journal} {Phys. Rev. Lett.}\ }\textbf {\bibinfo {volume} {119}},\
  \bibinfo {pages} {200402} (\bibinfo {year} {2017})}\BibitemShut {NoStop}%
\bibitem [{\citenamefont {D'Alessio}\ and\ \citenamefont
  {Rigol}(2014)}]{Dalessio2014}%
  \BibitemOpen
  \bibfield  {author} {\bibinfo {author} {\bibfnamefont {L.}~\bibnamefont
  {D'Alessio}}\ and\ \bibinfo {author} {\bibfnamefont {M.}~\bibnamefont
  {Rigol}},\ }\href {https://doi.org/10.1103%2Fphysrevx.4.041048} {\bibfield
  {journal} {\bibinfo  {journal} {Phys. Rev. X}\ }\textbf {\bibinfo {volume}
  {4}},\ \bibinfo {pages} {041048} (\bibinfo {year} {2014})}\BibitemShut
  {NoStop}%
\bibitem [{\citenamefont {Ponte}\ \emph
  {et~al.}(2015{\natexlab{a}})\citenamefont {Ponte}, \citenamefont
  {Papi{\'{c}}}, \citenamefont {Huveneers},\ and\ \citenamefont
  {Abanin}}]{Ponte2015b}%
  \BibitemOpen
  \bibfield  {author} {\bibinfo {author} {\bibfnamefont {P.}~\bibnamefont
  {Ponte}}, \bibinfo {author} {\bibfnamefont {Z.}~\bibnamefont {Papi{\'{c}}}},
  \bibinfo {author} {\bibfnamefont {F.}~\bibnamefont {Huveneers}}, \ and\
  \bibinfo {author} {\bibfnamefont {D.~A.}\ \bibnamefont {Abanin}},\ }\href
  {https://doi.org/10.1103%2Fphysrevlett.114.140401} {\bibfield  {journal}
  {\bibinfo  {journal} {Phys. Rev. Lett.}\ }\textbf {\bibinfo {volume} {114}},\
  \bibinfo {pages} {140401} (\bibinfo {year} {2015}{\natexlab{a}})}\BibitemShut
  {NoStop}%
\bibitem [{\citenamefont {Ponte}\ \emph
  {et~al.}(2015{\natexlab{b}})\citenamefont {Ponte}, \citenamefont {Chandran},
  \citenamefont {Papi{\'{c}}},\ and\ \citenamefont {Abanin}}]{Ponte2015}%
  \BibitemOpen
  \bibfield  {author} {\bibinfo {author} {\bibfnamefont {P.}~\bibnamefont
  {Ponte}}, \bibinfo {author} {\bibfnamefont {A.}~\bibnamefont {Chandran}},
  \bibinfo {author} {\bibfnamefont {Z.}~\bibnamefont {Papi{\'{c}}}}, \ and\
  \bibinfo {author} {\bibfnamefont {D.~A.}\ \bibnamefont {Abanin}},\ }\href
  {https://doi.org/10.1016%2Fj.aop.2014.11.008} {\bibfield  {journal} {\bibinfo
   {journal} {Ann. Phys.}\ }\textbf {\bibinfo {volume} {353}},\ \bibinfo
  {pages} {196} (\bibinfo {year} {2015}{\natexlab{b}})}\BibitemShut {NoStop}%
\bibitem [{\citenamefont {Lazarides}\ \emph {et~al.}(2015)\citenamefont
  {Lazarides}, \citenamefont {Das},\ and\ \citenamefont
  {Moessner}}]{Lazarides2015}%
  \BibitemOpen
  \bibfield  {author} {\bibinfo {author} {\bibfnamefont {A.}~\bibnamefont
  {Lazarides}}, \bibinfo {author} {\bibfnamefont {A.}~\bibnamefont {Das}}, \
  and\ \bibinfo {author} {\bibfnamefont {R.}~\bibnamefont {Moessner}},\ }\href
  {https://doi.org/10.1103%2Fphysrevlett.115.030402} {\bibfield  {journal}
  {\bibinfo  {journal} {Phys. Rev. Lett.}\ }\textbf {\bibinfo {volume} {115}},\
  \bibinfo {pages} {030402} (\bibinfo {year} {2015})}\BibitemShut {NoStop}%
\bibitem [{\citenamefont {Keser}\ \emph {et~al.}(2016)\citenamefont {Keser},
  \citenamefont {Ganeshan}, \citenamefont {Refael},\ and\ \citenamefont
  {Galitski}}]{Keser2016}%
  \BibitemOpen
  \bibfield  {author} {\bibinfo {author} {\bibfnamefont {A.~C.}\ \bibnamefont
  {Keser}}, \bibinfo {author} {\bibfnamefont {S.}~\bibnamefont {Ganeshan}},
  \bibinfo {author} {\bibfnamefont {G.}~\bibnamefont {Refael}}, \ and\ \bibinfo
  {author} {\bibfnamefont {V.}~\bibnamefont {Galitski}},\ }\href
  {https://doi.org/10.1103%2Fphysrevb.94.085120} {\bibfield  {journal}
  {\bibinfo  {journal} {Phys. Rev. B}\ }\textbf {\bibinfo {volume} {94}},\
  \bibinfo {pages} {085120} (\bibinfo {year} {2016})}\BibitemShut {NoStop}%
\bibitem [{\citenamefont {Bordia}\ \emph {et~al.}(2017)\citenamefont {Bordia},
  \citenamefont {Lüschen}, \citenamefont {Schneider}, \citenamefont {Knap},\
  and\ \citenamefont {Bloch}}]{Bordia2017}%
  \BibitemOpen
  \bibfield  {author} {\bibinfo {author} {\bibfnamefont {P.}~\bibnamefont
  {Bordia}}, \bibinfo {author} {\bibfnamefont {H.}~\bibnamefont {Lüschen}},
  \bibinfo {author} {\bibfnamefont {U.}~\bibnamefont {Schneider}}, \bibinfo
  {author} {\bibfnamefont {M.}~\bibnamefont {Knap}}, \ and\ \bibinfo {author}
  {\bibfnamefont {I.}~\bibnamefont {Bloch}},\ }\href {\doibase
  10.1038/nphys4020} {\bibfield  {journal} {\bibinfo  {journal} {Nat. Phys.}\
  }\textbf {\bibinfo {volume} {13}},\ \bibinfo {pages} {460} (\bibinfo {year}
  {2017})}\BibitemShut {NoStop}%
\bibitem [{\citenamefont {Notarnicola}\ \emph {et~al.}(2020)\citenamefont
  {Notarnicola}, \citenamefont {Silva}, \citenamefont {Fazio},\ and\
  \citenamefont {Russomanno}}]{Notarnicola2020}%
  \BibitemOpen
  \bibfield  {author} {\bibinfo {author} {\bibfnamefont {S.}~\bibnamefont
  {Notarnicola}}, \bibinfo {author} {\bibfnamefont {A.}~\bibnamefont {Silva}},
  \bibinfo {author} {\bibfnamefont {R.}~\bibnamefont {Fazio}}, \ and\ \bibinfo
  {author} {\bibfnamefont {A.}~\bibnamefont {Russomanno}},\ }\href
  {https://doi.org/10.1088%2F1742-5468%2Fab6de4} {\bibfield  {journal}
  {\bibinfo  {journal} {J. Stat. Mech.}\ }\textbf {\bibinfo {volume} {2020}},\
  \bibinfo {pages} {024008} (\bibinfo {year} {2020})}\BibitemShut {NoStop}%
\bibitem [{\citenamefont {Rylands}\ \emph {et~al.}(2020)\citenamefont
  {Rylands}, \citenamefont {Rozenbaum}, \citenamefont {Galitski},\ and\
  \citenamefont {Konik}}]{Rylands2020}%
  \BibitemOpen
  \bibfield  {author} {\bibinfo {author} {\bibfnamefont {C.}~\bibnamefont
  {Rylands}}, \bibinfo {author} {\bibfnamefont {E.~B.}\ \bibnamefont
  {Rozenbaum}}, \bibinfo {author} {\bibfnamefont {V.}~\bibnamefont {Galitski}},
  \ and\ \bibinfo {author} {\bibfnamefont {R.}~\bibnamefont {Konik}},\ }\href
  {https://doi.org/10.1103%2Fphysrevlett.124.155302} {\bibfield  {journal}
  {\bibinfo  {journal} {Phys. Rev. Lett.}\ }\textbf {\bibinfo {volume} {124}},\
  \bibinfo {pages} {155302} (\bibinfo {year} {2020})}\BibitemShut {NoStop}%
\bibitem [{\citenamefont {Vuatelet}\ and\ \citenamefont
  {Ran{\c{c}}on}(2021)}]{Vuatelet2021}%
  \BibitemOpen
  \bibfield  {author} {\bibinfo {author} {\bibfnamefont {V.}~\bibnamefont
  {Vuatelet}}\ and\ \bibinfo {author} {\bibfnamefont {A.}~\bibnamefont
  {Ran{\c{c}}on}},\ }\href {https://doi.org/10.1103%2Fphysreva.104.043302}
  {\bibfield  {journal} {\bibinfo  {journal} {Phys. Rev. A}\ }\textbf {\bibinfo
  {volume} {104}},\ \bibinfo {pages} {043302} (\bibinfo {year}
  {2021})}\BibitemShut {NoStop}%
\bibitem [{\citenamefont {Abanin}\ \emph {et~al.}(2015)\citenamefont {Abanin},
  \citenamefont {{De~Roeck}},\ and\ \citenamefont {Huveneers}}]{Abanin2015}%
  \BibitemOpen
  \bibfield  {author} {\bibinfo {author} {\bibfnamefont {D.~A.}\ \bibnamefont
  {Abanin}}, \bibinfo {author} {\bibfnamefont {W.}~\bibnamefont {{De~Roeck}}},
  \ and\ \bibinfo {author} {\bibfnamefont {F.}~\bibnamefont {Huveneers}},\
  }\href {https://doi.org/10.1103%2Fphysrevlett.115.256803} {\bibfield
  {journal} {\bibinfo  {journal} {Phys. Rev. Lett.}\ }\textbf {\bibinfo
  {volume} {115}},\ \bibinfo {pages} {256803} (\bibinfo {year}
  {2015})}\BibitemShut {NoStop}%
\bibitem [{\citenamefont {Bukov}\ \emph {et~al.}(2015)\citenamefont {Bukov},
  \citenamefont {Gopalakrishnan}, \citenamefont {Knap},\ and\ \citenamefont
  {Demler}}]{Bukov2015}%
  \BibitemOpen
  \bibfield  {author} {\bibinfo {author} {\bibfnamefont {M.}~\bibnamefont
  {Bukov}}, \bibinfo {author} {\bibfnamefont {S.}~\bibnamefont
  {Gopalakrishnan}}, \bibinfo {author} {\bibfnamefont {M.}~\bibnamefont
  {Knap}}, \ and\ \bibinfo {author} {\bibfnamefont {E.}~\bibnamefont
  {Demler}},\ }\href {https://doi.org/10.1103%2Fphysrevlett.115.205301}
  {\bibfield  {journal} {\bibinfo  {journal} {Phys. Rev. Lett.}\ }\textbf
  {\bibinfo {volume} {115}},\ \bibinfo {pages} {205301} (\bibinfo {year}
  {2015})}\BibitemShut {NoStop}%
\bibitem [{\citenamefont {Else}\ \emph {et~al.}(2017)\citenamefont {Else},
  \citenamefont {Bauer},\ and\ \citenamefont {Nayak}}]{Else2017}%
  \BibitemOpen
  \bibfield  {author} {\bibinfo {author} {\bibfnamefont {D.~V.}\ \bibnamefont
  {Else}}, \bibinfo {author} {\bibfnamefont {B.}~\bibnamefont {Bauer}}, \ and\
  \bibinfo {author} {\bibfnamefont {C.}~\bibnamefont {Nayak}},\ }\href
  {https://doi.org/10.1103%2Fphysrevx.7.011026} {\bibfield  {journal} {\bibinfo
   {journal} {Phys. Rev. X}\ }\textbf {\bibinfo {volume} {7}},\ \bibinfo
  {pages} {011026} (\bibinfo {year} {2017})}\BibitemShut {NoStop}%
\bibitem [{\citenamefont {Howell}\ \emph {et~al.}(2019)\citenamefont {Howell},
  \citenamefont {Weinberg}, \citenamefont {Sels}, \citenamefont {Polkovnikov},\
  and\ \citenamefont {Bukov}}]{Howell2019}%
  \BibitemOpen
  \bibfield  {author} {\bibinfo {author} {\bibfnamefont {O.}~\bibnamefont
  {Howell}}, \bibinfo {author} {\bibfnamefont {P.}~\bibnamefont {Weinberg}},
  \bibinfo {author} {\bibfnamefont {D.}~\bibnamefont {Sels}}, \bibinfo {author}
  {\bibfnamefont {A.}~\bibnamefont {Polkovnikov}}, \ and\ \bibinfo {author}
  {\bibfnamefont {M.}~\bibnamefont {Bukov}},\ }\href
  {https://doi.org/10.1103/PhysRevLett.122.010602} {\bibfield  {journal}
  {\bibinfo  {journal} {Phys. Rev. Lett.}\ }\textbf {\bibinfo {volume} {122}},\
  \bibinfo {pages} {010602} (\bibinfo {year} {2019})}\BibitemShut {NoStop}%
\bibitem [{\citenamefont {Rubio-Abadal}\ \emph {et~al.}(2020)\citenamefont
  {Rubio-Abadal}, \citenamefont {Ippoliti}, \citenamefont {Hollerith},
  \citenamefont {Wei}, \citenamefont {Rui}, \citenamefont {Sondhi},
  \citenamefont {Khemani}, \citenamefont {Gross},\ and\ \citenamefont
  {Bloch}}]{Abadal2020}%
  \BibitemOpen
  \bibfield  {author} {\bibinfo {author} {\bibfnamefont {A.}~\bibnamefont
  {Rubio-Abadal}}, \bibinfo {author} {\bibfnamefont {M.}~\bibnamefont
  {Ippoliti}}, \bibinfo {author} {\bibfnamefont {S.}~\bibnamefont {Hollerith}},
  \bibinfo {author} {\bibfnamefont {D.}~\bibnamefont {Wei}}, \bibinfo {author}
  {\bibfnamefont {J.}~\bibnamefont {Rui}}, \bibinfo {author} {\bibfnamefont
  {S.~L.}\ \bibnamefont {Sondhi}}, \bibinfo {author} {\bibfnamefont
  {V.}~\bibnamefont {Khemani}}, \bibinfo {author} {\bibfnamefont
  {C.}~\bibnamefont {Gross}}, \ and\ \bibinfo {author} {\bibfnamefont
  {I.}~\bibnamefont {Bloch}},\ }\href
  {https://doi.org/10.1103%2Fphysrevx.10.021044} {\bibfield  {journal}
  {\bibinfo  {journal} {Phys. Rev. X}\ }\textbf {\bibinfo {volume} {10}},\
  \bibinfo {pages} {021044} (\bibinfo {year} {2020})}\BibitemShut {NoStop}%
\bibitem [{\citenamefont {Hodson}\ and\ \citenamefont
  {Jarzynski}(2021)}]{Hodson2021}%
  \BibitemOpen
  \bibfield  {author} {\bibinfo {author} {\bibfnamefont {W.}~\bibnamefont
  {Hodson}}\ and\ \bibinfo {author} {\bibfnamefont {C.}~\bibnamefont
  {Jarzynski}},\ }\href {https://doi.org/10.1103/PhysRevResearch.3.013219}
  {\bibfield  {journal} {\bibinfo  {journal} {Phys. Rev. Research}\ }\textbf
  {\bibinfo {volume} {3}},\ \bibinfo {pages} {013219} (\bibinfo {year}
  {2021})}\BibitemShut {NoStop}%
\bibitem [{\citenamefont {Bhakuni}\ \emph {et~al.}(2021)\citenamefont
  {Bhakuni}, \citenamefont {Santos},\ and\ \citenamefont {Lev}}]{Bhakuni2021}%
  \BibitemOpen
  \bibfield  {author} {\bibinfo {author} {\bibfnamefont {D.~S.}\ \bibnamefont
  {Bhakuni}}, \bibinfo {author} {\bibfnamefont {L.~F.}\ \bibnamefont {Santos}},
  \ and\ \bibinfo {author} {\bibfnamefont {Y.~B.}\ \bibnamefont {Lev}},\ }\href
  {\doibase 10.1103/PhysRevB.104.L140301} {\bibfield  {journal} {\bibinfo
  {journal} {Phys. Rev. B}\ }\textbf {\bibinfo {volume} {104}},\ \bibinfo
  {pages} {L140301} (\bibinfo {year} {2021})}\BibitemShut {NoStop}%
\bibitem [{\citenamefont {Casati}\ \emph {et~al.}(1979)\citenamefont {Casati},
  \citenamefont {Chirikov}, \citenamefont {Izraelev},\ and\ \citenamefont
  {Ford}}]{Casati1979}%
  \BibitemOpen
  \bibfield  {author} {\bibinfo {author} {\bibfnamefont {G.}~\bibnamefont
  {Casati}}, \bibinfo {author} {\bibfnamefont {B.~V.}\ \bibnamefont
  {Chirikov}}, \bibinfo {author} {\bibfnamefont {F.~M.}\ \bibnamefont
  {Izraelev}}, \ and\ \bibinfo {author} {\bibfnamefont {J.}~\bibnamefont
  {Ford}},\ }in\ \href {https://doi.org/10.1007/BFb0021757} {\emph {\bibinfo
  {booktitle} {Stochastic Behavior in Classical and Quantum Hamiltonian
  Systems}}},\ \bibinfo {editor} {edited by\ \bibinfo {editor} {\bibfnamefont
  {G.}~\bibnamefont {Casati}}\ and\ \bibinfo {editor} {\bibfnamefont
  {J.}~\bibnamefont {Ford}}}\ (\bibinfo  {publisher} {Springer Berlin
  Heidelberg},\ \bibinfo {address} {Berlin, Heidelberg},\ \bibinfo {year}
  {1979})\ pp.\ \bibinfo {pages} {334--352}\BibitemShut {NoStop}%
\bibitem [{\citenamefont {Moore}\ \emph {et~al.}(1995)\citenamefont {Moore},
  \citenamefont {Robinson}, \citenamefont {Bharucha}, \citenamefont
  {Sundaram},\ and\ \citenamefont {Raizen}}]{Moore95}%
  \BibitemOpen
  \bibfield  {author} {\bibinfo {author} {\bibfnamefont {F.~L.}\ \bibnamefont
  {Moore}}, \bibinfo {author} {\bibfnamefont {J.~C.}\ \bibnamefont {Robinson}},
  \bibinfo {author} {\bibfnamefont {C.~F.}\ \bibnamefont {Bharucha}}, \bibinfo
  {author} {\bibfnamefont {B.}~\bibnamefont {Sundaram}}, \ and\ \bibinfo
  {author} {\bibfnamefont {M.~G.}\ \bibnamefont {Raizen}},\ }\href
  {https://doi.org/10.1103/PhysRevLett.75.4598} {\bibfield  {journal} {\bibinfo
   {journal} {Phys. Rev. Lett.}\ }\textbf {\bibinfo {volume} {75}},\ \bibinfo
  {pages} {4598} (\bibinfo {year} {1995})}\BibitemShut {NoStop}%
\bibitem [{\citenamefont {Chab\'e}\ \emph {et~al.}(2008)\citenamefont
  {Chab\'e}, \citenamefont {Lemari\'e}, \citenamefont {Gr\'emaud},
  \citenamefont {Delande}, \citenamefont {Szriftgiser},\ and\ \citenamefont
  {Garreau}}]{Chabe2008}%
  \BibitemOpen
  \bibfield  {author} {\bibinfo {author} {\bibfnamefont {J.}~\bibnamefont
  {Chab\'e}}, \bibinfo {author} {\bibfnamefont {G.}~\bibnamefont {Lemari\'e}},
  \bibinfo {author} {\bibfnamefont {B.}~\bibnamefont {Gr\'emaud}}, \bibinfo
  {author} {\bibfnamefont {D.}~\bibnamefont {Delande}}, \bibinfo {author}
  {\bibfnamefont {P.}~\bibnamefont {Szriftgiser}}, \ and\ \bibinfo {author}
  {\bibfnamefont {J.~C.}\ \bibnamefont {Garreau}},\ }\href
  {https://doi.org/10.1103/PhysRevLett.101.255702} {\bibfield  {journal}
  {\bibinfo  {journal} {Phys. Rev. Lett.}\ }\textbf {\bibinfo {volume} {101}},\
  \bibinfo {pages} {255702} (\bibinfo {year} {2008})}\BibitemShut {NoStop}%
\bibitem [{\citenamefont {Hainaut}\ \emph {et~al.}(2017)\citenamefont
  {Hainaut}, \citenamefont {Manai}, \citenamefont {Chicireanu}, \citenamefont
  {Cl\'ement}, \citenamefont {Zemmouri}, \citenamefont {Garreau}, \citenamefont
  {Szriftgiser}, \citenamefont {Lemari\'e}, \citenamefont {Cherroret},\ and\
  \citenamefont {Delande}}]{Hainaut17}%
  \BibitemOpen
  \bibfield  {author} {\bibinfo {author} {\bibfnamefont {C.}~\bibnamefont
  {Hainaut}}, \bibinfo {author} {\bibfnamefont {I.}~\bibnamefont {Manai}},
  \bibinfo {author} {\bibfnamefont {R.}~\bibnamefont {Chicireanu}}, \bibinfo
  {author} {\bibfnamefont {J.-F.}\ \bibnamefont {Cl\'ement}}, \bibinfo {author}
  {\bibfnamefont {S.}~\bibnamefont {Zemmouri}}, \bibinfo {author}
  {\bibfnamefont {J.~C.}\ \bibnamefont {Garreau}}, \bibinfo {author}
  {\bibfnamefont {P.}~\bibnamefont {Szriftgiser}}, \bibinfo {author}
  {\bibfnamefont {G.}~\bibnamefont {Lemari\'e}}, \bibinfo {author}
  {\bibfnamefont {N.}~\bibnamefont {Cherroret}}, \ and\ \bibinfo {author}
  {\bibfnamefont {D.}~\bibnamefont {Delande}},\ }\href
  {https://doi.org/10.1103/PhysRevLett.118.184101} {\bibfield  {journal}
  {\bibinfo  {journal} {Phys. Rev. Lett.}\ }\textbf {\bibinfo {volume} {118}},\
  \bibinfo {pages} {184101} (\bibinfo {year} {2017})}\BibitemShut {NoStop}%
\bibitem [{\citenamefont {Hainaut}\ \emph {et~al.}(2018)\citenamefont
  {Hainaut}, \citenamefont {Manai}, \citenamefont {Cl{\'{e}}ment},
  \citenamefont {Garreau}, \citenamefont {Szriftgiser}, \citenamefont
  {Lemari{\'{e}}}, \citenamefont {Cherroret}, \citenamefont {Delande},\ and\
  \citenamefont {Chicireanu}}]{Hainaut18}%
  \BibitemOpen
  \bibfield  {author} {\bibinfo {author} {\bibfnamefont {C.}~\bibnamefont
  {Hainaut}}, \bibinfo {author} {\bibfnamefont {I.}~\bibnamefont {Manai}},
  \bibinfo {author} {\bibfnamefont {J.-F.}\ \bibnamefont {Cl{\'{e}}ment}},
  \bibinfo {author} {\bibfnamefont {J.~C.}\ \bibnamefont {Garreau}}, \bibinfo
  {author} {\bibfnamefont {P.}~\bibnamefont {Szriftgiser}}, \bibinfo {author}
  {\bibfnamefont {G.}~\bibnamefont {Lemari{\'{e}}}}, \bibinfo {author}
  {\bibfnamefont {N.}~\bibnamefont {Cherroret}}, \bibinfo {author}
  {\bibfnamefont {D.}~\bibnamefont {Delande}}, \ and\ \bibinfo {author}
  {\bibfnamefont {R.}~\bibnamefont {Chicireanu}},\ }\href
  {https://doi.org/10.1038%2Fs41467-018-03481-9} {\bibfield  {journal}
  {\bibinfo  {journal} {Nat. Commun.}\ }\textbf {\bibinfo {volume} {9}},\
  \bibinfo {pages} {1382} (\bibinfo {year} {2018})}\BibitemShut {NoStop}%
\bibitem [{\citenamefont {Hainaut}\ \emph {et~al.}(2022)\citenamefont
  {Hainaut}, \citenamefont {Cl{\'{e}}ment}, \citenamefont {Szriftgiser},
  \citenamefont {Garreau}, \citenamefont {Ran{\c{c}}on},\ and\ \citenamefont
  {Chicireanu}}]{Hainaut2022}%
  \BibitemOpen
  \bibfield  {author} {\bibinfo {author} {\bibfnamefont {C.}~\bibnamefont
  {Hainaut}}, \bibinfo {author} {\bibfnamefont {J.-F.}\ \bibnamefont
  {Cl{\'{e}}ment}}, \bibinfo {author} {\bibfnamefont {P.}~\bibnamefont
  {Szriftgiser}}, \bibinfo {author} {\bibfnamefont {J.~C.}\ \bibnamefont
  {Garreau}}, \bibinfo {author} {\bibfnamefont {A.}~\bibnamefont
  {Ran{\c{c}}on}}, \ and\ \bibinfo {author} {\bibfnamefont {R.}~\bibnamefont
  {Chicireanu}},\ }\href {https://doi.org/10.1140%2Fepjd%2Fs10053-022-00426-2}
  {\bibfield  {journal} {\bibinfo  {journal} {Eur. Phys. J. D}\ }\textbf
  {\bibinfo {volume} {76}},\ \bibinfo {pages} {103} (\bibinfo {year}
  {2022})}\BibitemShut {NoStop}%
\bibitem [{\citenamefont {Shepelyansky}(1993)}]{Shepelyansky1993}%
  \BibitemOpen
  \bibfield  {author} {\bibinfo {author} {\bibfnamefont {D.~L.}\ \bibnamefont
  {Shepelyansky}},\ }\href {https://doi.org/10.1103/PhysRevLett.70.1787}
  {\bibfield  {journal} {\bibinfo  {journal} {Phys. Rev. Lett.}\ }\textbf
  {\bibinfo {volume} {70}},\ \bibinfo {pages} {1787} (\bibinfo {year}
  {1993})}\BibitemShut {NoStop}%
\bibitem [{\citenamefont {Gligori{\'{c}}}\ \emph {et~al.}(2011)\citenamefont
  {Gligori{\'{c}}}, \citenamefont {Bodyfelt},\ and\ \citenamefont
  {Flach}}]{Gligoric2011}%
  \BibitemOpen
  \bibfield  {author} {\bibinfo {author} {\bibfnamefont {G.}~\bibnamefont
  {Gligori{\'{c}}}}, \bibinfo {author} {\bibfnamefont {J.~D.}\ \bibnamefont
  {Bodyfelt}}, \ and\ \bibinfo {author} {\bibfnamefont {S.}~\bibnamefont
  {Flach}},\ }\href {https://doi.org/10.1209%2F0295-5075%2F96%2F30004}
  {\bibfield  {journal} {\bibinfo  {journal} {EPL}\ }\textbf {\bibinfo {volume}
  {96}},\ \bibinfo {pages} {30004} (\bibinfo {year} {2011})}\BibitemShut
  {NoStop}%
\bibitem [{\citenamefont {Cherroret}\ \emph {et~al.}(2014)\citenamefont
  {Cherroret}, \citenamefont {Vermersch}, \citenamefont {Garreau},\ and\
  \citenamefont {Delande}}]{Cherroret14}%
  \BibitemOpen
  \bibfield  {author} {\bibinfo {author} {\bibfnamefont {N.}~\bibnamefont
  {Cherroret}}, \bibinfo {author} {\bibfnamefont {B.}~\bibnamefont
  {Vermersch}}, \bibinfo {author} {\bibfnamefont {J.~C.}\ \bibnamefont
  {Garreau}}, \ and\ \bibinfo {author} {\bibfnamefont {D.}~\bibnamefont
  {Delande}},\ }\href {https://doi.org/10.1103%2Fphysrevlett.112.170603}
  {\bibfield  {journal} {\bibinfo  {journal} {Phys. Rev. Lett.}\ }\textbf
  {\bibinfo {volume} {112}},\ \bibinfo {pages} {170603} (\bibinfo {year}
  {2014})}\BibitemShut {NoStop}%
\bibitem [{\citenamefont {Cherroret}(2016)}]{Cherroret16}%
  \BibitemOpen
  \bibfield  {author} {\bibinfo {author} {\bibfnamefont {N.}~\bibnamefont
  {Cherroret}},\ }\href {https://doi.org/10.1088%2F0953-8984%2F29%2F2%2F024002}
  {\bibfield  {journal} {\bibinfo  {journal} {J. Phys.: Condens. Matter}\
  }\textbf {\bibinfo {volume} {29}},\ \bibinfo {pages} {024002} (\bibinfo
  {year} {2016})}\BibitemShut {NoStop}%
\bibitem [{\citenamefont {Lellouch}\ \emph {et~al.}(2020)\citenamefont
  {Lellouch}, \citenamefont {Ran{\c{c}}on}, \citenamefont {{De~Bi{\`{e}}vre}},
  \citenamefont {Delande},\ and\ \citenamefont {Garreau}}]{Lellouch2020}%
  \BibitemOpen
  \bibfield  {author} {\bibinfo {author} {\bibfnamefont {S.}~\bibnamefont
  {Lellouch}}, \bibinfo {author} {\bibfnamefont {A.}~\bibnamefont
  {Ran{\c{c}}on}}, \bibinfo {author} {\bibfnamefont {S.}~\bibnamefont
  {{De~Bi{\`{e}}vre}}}, \bibinfo {author} {\bibfnamefont {D.}~\bibnamefont
  {Delande}}, \ and\ \bibinfo {author} {\bibfnamefont {J.~C.}\ \bibnamefont
  {Garreau}},\ }\href {https://doi.org/10.1103%2Fphysreva.101.043624}
  {\bibfield  {journal} {\bibinfo  {journal} {Phys. Rev. A}\ }\textbf {\bibinfo
  {volume} {101}},\ \bibinfo {pages} {043624} (\bibinfo {year}
  {2020})}\BibitemShut {NoStop}%
\bibitem [{\citenamefont {Haldar}\ \emph {et~al.}(2021)\citenamefont {Haldar},
  \citenamefont {Mu}, \citenamefont {Georgeot}, \citenamefont {Gong},
  \citenamefont {Miniatura},\ and\ \citenamefont {Lemarié}}]{Haldar2022}%
  \BibitemOpen
  \bibfield  {author} {\bibinfo {author} {\bibfnamefont {P.}~\bibnamefont
  {Haldar}}, \bibinfo {author} {\bibfnamefont {S.}~\bibnamefont {Mu}}, \bibinfo
  {author} {\bibfnamefont {B.}~\bibnamefont {Georgeot}}, \bibinfo {author}
  {\bibfnamefont {J.}~\bibnamefont {Gong}}, \bibinfo {author} {\bibfnamefont
  {C.}~\bibnamefont {Miniatura}}, \ and\ \bibinfo {author} {\bibfnamefont
  {G.}~\bibnamefont {Lemarié}},\ }\href {https://arxiv.org/abs/2109.14347}
  {\bibfield  {journal} {\bibinfo  {journal} {arXiv:2109.14347}\ } (\bibinfo
  {year} {2021})}\BibitemShut {NoStop}%
\bibitem [{\citenamefont {Mu}\ \emph {et~al.}(2022)\citenamefont {Mu},
  \citenamefont {Macé}, \citenamefont {Gong}, \citenamefont {Miniatura},
  \citenamefont {Lemarié},\ and\ \citenamefont {Albert}}]{Mu2022}%
  \BibitemOpen
  \bibfield  {author} {\bibinfo {author} {\bibfnamefont {S.}~\bibnamefont
  {Mu}}, \bibinfo {author} {\bibfnamefont {N.}~\bibnamefont {Macé}}, \bibinfo
  {author} {\bibfnamefont {J.}~\bibnamefont {Gong}}, \bibinfo {author}
  {\bibfnamefont {C.}~\bibnamefont {Miniatura}}, \bibinfo {author}
  {\bibfnamefont {G.}~\bibnamefont {Lemarié}}, \ and\ \bibinfo {author}
  {\bibfnamefont {M.}~\bibnamefont {Albert}},\ }\href
  {https://arxiv.org/abs/2207.06951} {\bibfield  {journal} {\bibinfo  {journal}
  {arXiv:2207.06951}\ } (\bibinfo {year} {2022})}\BibitemShut {NoStop}%
\bibitem [{\citenamefont {Scoquart}\ \emph
  {et~al.}(2020{\natexlab{a}})\citenamefont {Scoquart}, \citenamefont
  {Larr{\'{e}}}, \citenamefont {Delande},\ and\ \citenamefont
  {Cherroret}}]{Scoquart2020b}%
  \BibitemOpen
  \bibfield  {author} {\bibinfo {author} {\bibfnamefont {T.}~\bibnamefont
  {Scoquart}}, \bibinfo {author} {\bibfnamefont {P.-{\'{E}}.}\ \bibnamefont
  {Larr{\'{e}}}}, \bibinfo {author} {\bibfnamefont {D.}~\bibnamefont
  {Delande}}, \ and\ \bibinfo {author} {\bibfnamefont {N.}~\bibnamefont
  {Cherroret}},\ }\href {https://doi.org/10.1209%2F0295-5075%2F132%2F66001}
  {\bibfield  {journal} {\bibinfo  {journal} {EPL}\ }\textbf {\bibinfo {volume}
  {132}},\ \bibinfo {pages} {66001} (\bibinfo {year}
  {2020}{\natexlab{a}})}\BibitemShut {NoStop}%
\bibitem [{\citenamefont {Cherroret}\ \emph {et~al.}(2021)\citenamefont
  {Cherroret}, \citenamefont {Scoquart},\ and\ \citenamefont
  {Delande}}]{Cherroret2021}%
  \BibitemOpen
  \bibfield  {author} {\bibinfo {author} {\bibfnamefont {N.}~\bibnamefont
  {Cherroret}}, \bibinfo {author} {\bibfnamefont {T.}~\bibnamefont {Scoquart}},
  \ and\ \bibinfo {author} {\bibfnamefont {D.}~\bibnamefont {Delande}},\ }\href
  {https://doi.org/10.1016%2Fj.aop.2021.168543} {\bibfield  {journal} {\bibinfo
   {journal} {Ann. Phys.}\ }\textbf {\bibinfo {volume} {435}},\ \bibinfo
  {pages} {168543} (\bibinfo {year} {2021})}\BibitemShut {NoStop}%
\bibitem [{Note1()}]{Note1}%
  \BibitemOpen
  \bibinfo {note} {Note that the pre-thermalization scenario studied in
  Sec.~\ref {Sec:Pre-thermal dynamics} occurs regardless the structure of the
  interaction term in the nonlinear wave equation (\ref {eq:GPE}), be it local
  or nonlocal, in the latter case at least provided the response function
  varies rapidly at the scale of the healing length.}\BibitemShut {Stop}%
\bibitem [{\citenamefont {Fishman}\ \emph {et~al.}(1982)\citenamefont
  {Fishman}, \citenamefont {Grempel},\ and\ \citenamefont
  {Prange}}]{Fishman1982}%
  \BibitemOpen
  \bibfield  {author} {\bibinfo {author} {\bibfnamefont {S.}~\bibnamefont
  {Fishman}}, \bibinfo {author} {\bibfnamefont {D.~R.}\ \bibnamefont
  {Grempel}}, \ and\ \bibinfo {author} {\bibfnamefont {R.~E.}\ \bibnamefont
  {Prange}},\ }\href {https://doi.org/10.1103/PhysRevLett.49.509} {\bibfield
  {journal} {\bibinfo  {journal} {Phys. Rev. Lett.}\ }\textbf {\bibinfo
  {volume} {49}},\ \bibinfo {pages} {509} (\bibinfo {year} {1982})}\BibitemShut
  {NoStop}%
\bibitem [{\citenamefont {Grempel}\ \emph {et~al.}(1984)\citenamefont
  {Grempel}, \citenamefont {Prange},\ and\ \citenamefont
  {Fishman}}]{Grempel84}%
  \BibitemOpen
  \bibfield  {author} {\bibinfo {author} {\bibfnamefont {D.~R.}\ \bibnamefont
  {Grempel}}, \bibinfo {author} {\bibfnamefont {R.~E.}\ \bibnamefont {Prange}},
  \ and\ \bibinfo {author} {\bibfnamefont {S.}~\bibnamefont {Fishman}},\ }\href
  {https://doi.org/10.1103/PhysRevA.29.1639} {\bibfield  {journal} {\bibinfo
  {journal} {Phys. Rev. A}\ }\textbf {\bibinfo {volume} {29}},\ \bibinfo
  {pages} {1639} (\bibinfo {year} {1984})}\BibitemShut {NoStop}%
\bibitem [{\citenamefont {Shepelyansky}(1986)}]{Shepelyansky1986}%
  \BibitemOpen
  \bibfield  {author} {\bibinfo {author} {\bibfnamefont {D.~L.}\ \bibnamefont
  {Shepelyansky}},\ }\href {https://doi.org/10.1103/PhysRevLett.56.677}
  {\bibfield  {journal} {\bibinfo  {journal} {Phys. Rev. Lett.}\ }\textbf
  {\bibinfo {volume} {56}},\ \bibinfo {pages} {677} (\bibinfo {year}
  {1986})}\BibitemShut {NoStop}%
\bibitem [{\citenamefont {Izrailev}\ and\ \citenamefont
  {Shepelyanskii}(1980)}]{Izrailev1980}%
  \BibitemOpen
  \bibfield  {author} {\bibinfo {author} {\bibfnamefont {F.~M.}\ \bibnamefont
  {Izrailev}}\ and\ \bibinfo {author} {\bibfnamefont {D.~L.}\ \bibnamefont
  {Shepelyanskii}},\ }\href {https://doi.org/10.1007/BF01029131} {\bibfield
  {journal} {\bibinfo  {journal} {Theor. Math. Phys.}\ }\textbf {\bibinfo
  {volume} {43}},\ \bibinfo {pages} {553} (\bibinfo {year} {1980})}\BibitemShut
  {NoStop}%
\bibitem [{\citenamefont {Sokolov}\ \emph {et~al.}(2000)\citenamefont
  {Sokolov}, \citenamefont {Zhirov}, \citenamefont {Alonso},\ and\
  \citenamefont {Casati}}]{Sokolov2000}%
  \BibitemOpen
  \bibfield  {author} {\bibinfo {author} {\bibfnamefont {V.~V.}\ \bibnamefont
  {Sokolov}}, \bibinfo {author} {\bibfnamefont {O.~V.}\ \bibnamefont {Zhirov}},
  \bibinfo {author} {\bibfnamefont {D.}~\bibnamefont {Alonso}}, \ and\ \bibinfo
  {author} {\bibfnamefont {G.}~\bibnamefont {Casati}},\ }\href
  {https://doi.org/10.1103/PhysRevLett.84.3566} {\bibfield  {journal} {\bibinfo
   {journal} {Phys. Rev. Lett.}\ }\textbf {\bibinfo {volume} {84}},\ \bibinfo
  {pages} {3566} (\bibinfo {year} {2000})}\BibitemShut {NoStop}%
\bibitem [{\citenamefont {Wimberger}\ \emph {et~al.}(2004)\citenamefont
  {Wimberger}, \citenamefont {Guarneri},\ and\ \citenamefont
  {Fishman}}]{Wimberger2004}%
  \BibitemOpen
  \bibfield  {author} {\bibinfo {author} {\bibfnamefont {S.}~\bibnamefont
  {Wimberger}}, \bibinfo {author} {\bibfnamefont {I.}~\bibnamefont {Guarneri}},
  \ and\ \bibinfo {author} {\bibfnamefont {S.}~\bibnamefont {Fishman}},\ }\href
  {https://doi.org/10.1103/PhysRevLett.92.084102} {\bibfield  {journal}
  {\bibinfo  {journal} {Phys. Rev. Lett.}\ }\textbf {\bibinfo {volume} {92}},\
  \bibinfo {pages} {084102} (\bibinfo {year} {2004})}\BibitemShut {NoStop}%
\bibitem [{\citenamefont {Lepers}\ \emph {et~al.}(2008)\citenamefont {Lepers},
  \citenamefont {Zehnl\'e},\ and\ \citenamefont {Garreau}}]{Lepers2008}%
  \BibitemOpen
  \bibfield  {author} {\bibinfo {author} {\bibfnamefont {M.}~\bibnamefont
  {Lepers}}, \bibinfo {author} {\bibfnamefont {V.}~\bibnamefont {Zehnl\'e}}, \
  and\ \bibinfo {author} {\bibfnamefont {J.~C.}\ \bibnamefont {Garreau}},\
  }\href {https://doi.org/10.1103/PhysRevA.77.043628} {\bibfield  {journal}
  {\bibinfo  {journal} {Phys. Rev. A}\ }\textbf {\bibinfo {volume} {77}},\
  \bibinfo {pages} {043628} (\bibinfo {year} {2008})}\BibitemShut {NoStop}%
\bibitem [{\citenamefont {Cherroret}\ \emph {et~al.}(2015)\citenamefont
  {Cherroret}, \citenamefont {Karpiuk}, \citenamefont {Gr{\'{e}}maud},\ and\
  \citenamefont {Miniatura}}]{Cherroret2015}%
  \BibitemOpen
  \bibfield  {author} {\bibinfo {author} {\bibfnamefont {N.}~\bibnamefont
  {Cherroret}}, \bibinfo {author} {\bibfnamefont {T.}~\bibnamefont {Karpiuk}},
  \bibinfo {author} {\bibfnamefont {B.}~\bibnamefont {Gr{\'{e}}maud}}, \ and\
  \bibinfo {author} {\bibfnamefont {C.}~\bibnamefont {Miniatura}},\ }\href
  {https://doi.org/10.1103%2Fphysreva.92.063614} {\bibfield  {journal}
  {\bibinfo  {journal} {Phys. Rev. A}\ }\textbf {\bibinfo {volume} {92}},\
  \bibinfo {pages} {063614} (\bibinfo {year} {2015})}\BibitemShut {NoStop}%
\bibitem [{\citenamefont {Scoquart}\ \emph
  {et~al.}(2020{\natexlab{b}})\citenamefont {Scoquart}, \citenamefont
  {Wellens}, \citenamefont {Delande},\ and\ \citenamefont
  {Cherroret}}]{Scoquart2020}%
  \BibitemOpen
  \bibfield  {author} {\bibinfo {author} {\bibfnamefont {T.}~\bibnamefont
  {Scoquart}}, \bibinfo {author} {\bibfnamefont {T.}~\bibnamefont {Wellens}},
  \bibinfo {author} {\bibfnamefont {D.}~\bibnamefont {Delande}}, \ and\
  \bibinfo {author} {\bibfnamefont {N.}~\bibnamefont {Cherroret}},\ }\href
  {https://doi.org/10.1103%2Fphysrevresearch.2.033349} {\bibfield  {journal}
  {\bibinfo  {journal} {Phys. Rev. Research}\ }\textbf {\bibinfo {volume}
  {2}},\ \bibinfo {pages} {033349} (\bibinfo {year}
  {2020}{\natexlab{b}})}\BibitemShut {NoStop}%
\bibitem [{\citenamefont {Popov}(1972)}]{Popov1972}%
  \BibitemOpen
  \bibfield  {author} {\bibinfo {author} {\bibfnamefont {V.~N.}\ \bibnamefont
  {Popov}},\ }\href {https://link.springer.com/article/10.1007/BF01028373}
  {\bibfield  {journal} {\bibinfo  {journal} {Theor. Math. Phys.}\ }\textbf
  {\bibinfo {volume} {11}},\ \bibinfo {pages} {565} (\bibinfo {year}
  {1972})}\BibitemShut {NoStop}%
\bibitem [{\citenamefont {Popov}(1983)}]{Popov1983}%
  \BibitemOpen
  \bibfield  {author} {\bibinfo {author} {\bibfnamefont {V.~N.}\ \bibnamefont
  {Popov}},\ }\href@noop {} {\emph {\bibinfo {title} {Functional Integrals in
  Quantum Field Theory and Statistical Physics}}}\ (\bibinfo  {publisher}
  {Reidel},\ \bibinfo {address} {Dordrecht},\ \bibinfo {year}
  {1983})\BibitemShut {NoStop}%
\bibitem [{\citenamefont {Mora}\ and\ \citenamefont {Castin}(2003)}]{Mora2003}%
  \BibitemOpen
  \bibfield  {author} {\bibinfo {author} {\bibfnamefont {C.}~\bibnamefont
  {Mora}}\ and\ \bibinfo {author} {\bibfnamefont {Y.}~\bibnamefont {Castin}},\
  }\href {https://doi.org/10.1103/PhysRevA.67.053615} {\bibfield  {journal}
  {\bibinfo  {journal} {Phys. Rev. A}\ }\textbf {\bibinfo {volume} {67}},\
  \bibinfo {pages} {053615} (\bibinfo {year} {2003})}\BibitemShut {NoStop}%
\bibitem [{\citenamefont {Petrov}(2003)}]{Petrov2003}%
  \BibitemOpen
  \bibfield  {author} {\bibinfo {author} {\bibfnamefont {D.~S.}\ \bibnamefont
  {Petrov}},\ }\emph {\bibinfo {title} {Bose-Einstein Condensation in
  Low-Dimensional Trapped Gases}},\ \href@noop {} {Ph.D. thesis},\ \bibinfo
  {school} {University of Amsterdam} (\bibinfo {year} {2003})\BibitemShut
  {NoStop}%
\bibitem [{Note2()}]{Note2}%
  \BibitemOpen
  \bibinfo {note} {The way the subleading density fluctuations $\delta \rho
  (k,t)$ contribute to the $g_{1}$ function may be found in, e.g., Ref.~\cite
  {Larre2018} [see Eq.~(84)]}\BibitemShut {NoStop}%
\bibitem [{\citenamefont {Leboeuf}\ and\ \citenamefont
  {Pavloff}(2001)}]{Leboeuf2001}%
  \BibitemOpen
  \bibfield  {author} {\bibinfo {author} {\bibfnamefont {P.}~\bibnamefont
  {Leboeuf}}\ and\ \bibinfo {author} {\bibfnamefont {N.}~\bibnamefont
  {Pavloff}},\ }\href {https://doi.org/10.1103/PhysRevA.64.033602} {\bibfield
  {journal} {\bibinfo  {journal} {Phys. Rev. A}\ }\textbf {\bibinfo {volume}
  {64}},\ \bibinfo {pages} {033602} (\bibinfo {year} {2001})}\BibitemShut
  {NoStop}%
\bibitem [{\citenamefont {Larr{\'e}}\ \emph {et~al.}(2012)\citenamefont
  {Larr{\'e}}, \citenamefont {Pavloff},\ and\ \citenamefont
  {Kamchatnov}}]{Larre2012}%
  \BibitemOpen
  \bibfield  {author} {\bibinfo {author} {\bibfnamefont {P.-{\'E}.}\
  \bibnamefont {Larr{\'e}}}, \bibinfo {author} {\bibfnamefont {N.}~\bibnamefont
  {Pavloff}}, \ and\ \bibinfo {author} {\bibfnamefont {A.~M.}\ \bibnamefont
  {Kamchatnov}},\ }\href {https://doi.org/10.1103/PhysRevB.86.165304}
  {\bibfield  {journal} {\bibinfo  {journal} {Phys. Rev. B}\ }\textbf {\bibinfo
  {volume} {86}},\ \bibinfo {pages} {165304} (\bibinfo {year}
  {2012})}\BibitemShut {NoStop}%
\bibitem [{\citenamefont {Connaughton}\ \emph {et~al.}(2005)\citenamefont
  {Connaughton}, \citenamefont {Josserand}, \citenamefont {Picozzi},
  \citenamefont {Pomeau},\ and\ \citenamefont {Rica}}]{Connaughton2005}%
  \BibitemOpen
  \bibfield  {author} {\bibinfo {author} {\bibfnamefont {C.}~\bibnamefont
  {Connaughton}}, \bibinfo {author} {\bibfnamefont {C.}~\bibnamefont
  {Josserand}}, \bibinfo {author} {\bibfnamefont {A.}~\bibnamefont {Picozzi}},
  \bibinfo {author} {\bibfnamefont {Y.}~\bibnamefont {Pomeau}}, \ and\ \bibinfo
  {author} {\bibfnamefont {S.}~\bibnamefont {Rica}},\ }\href
  {https://doi.org/10.1103%2Fphysrevlett.95.263901} {\bibfield  {journal}
  {\bibinfo  {journal} {Phys. Rev. Lett.}\ }\textbf {\bibinfo {volume} {95}},\
  \bibinfo {pages} {263901} (\bibinfo {year} {2005})}\BibitemShut {NoStop}%
\bibitem [{\citenamefont {Landau}(1941{\natexlab{a}})}]{Landau1941a}%
  \BibitemOpen
  \bibfield  {author} {\bibinfo {author} {\bibfnamefont {L.~D.}\ \bibnamefont
  {Landau}},\ }\href
  {https://journals.aps.org/pr/abstract/10.1103/PhysRev.60.356} {\bibfield
  {journal} {\bibinfo  {journal} {Phys. Rev.}\ }\textbf {\bibinfo {volume}
  {60}},\ \bibinfo {pages} {356} (\bibinfo {year}
  {1941}{\natexlab{a}})}\BibitemShut {NoStop}%
\bibitem [{\citenamefont {Landau}(1941{\natexlab{b}})}]{Landau1941b}%
  \BibitemOpen
  \bibfield  {author} {\bibinfo {author} {\bibfnamefont {L.~D.}\ \bibnamefont
  {Landau}},\ }\href@noop {} {\bibfield  {journal} {\bibinfo  {journal} {J.
  Phys. USSR}\ }\textbf {\bibinfo {volume} {5}},\ \bibinfo {pages} {71}
  (\bibinfo {year} {1941}{\natexlab{b}})}\BibitemShut {NoStop}%
\bibitem [{\citenamefont {Leggett}(1999)}]{Leggett1999}%
  \BibitemOpen
  \bibfield  {author} {\bibinfo {author} {\bibfnamefont {A.~J.}\ \bibnamefont
  {Leggett}},\ }\href
  {https://journals.aps.org/rmp/abstract/10.1103/RevModPhys.71.S318} {\bibfield
   {journal} {\bibinfo  {journal} {Rev. Mod. Phys.}\ }\textbf {\bibinfo
  {volume} {71}},\ \bibinfo {pages} {S318} (\bibinfo {year}
  {1999})}\BibitemShut {NoStop}%
\bibitem [{\citenamefont {Pitaevskii}\ and\ \citenamefont
  {Stringari}(2016)}]{Pitaevskii2016}%
  \BibitemOpen
  \bibfield  {author} {\bibinfo {author} {\bibfnamefont {L.~P.}\ \bibnamefont
  {Pitaevskii}}\ and\ \bibinfo {author} {\bibfnamefont {S.}~\bibnamefont
  {Stringari}},\ }\href@noop {} {\emph {\bibinfo {title} {Bose-Einstein
  Condensation and Superfluidity}}}\ (\bibinfo  {publisher} {Oxford University
  Press},\ \bibinfo {address} {Oxford},\ \bibinfo {year} {2016})\BibitemShut
  {NoStop}%
\end{thebibliography}%

\end{document}